\documentclass[referee]{raa}            % referee version: for submission

\usepackage{graphicx,times}             %for PS/EPS graphics inclusion, new
\usepackage{natbib}
\usepackage{amssymb,amsmath}
\usepackage{color}

\bibpunct{(}{)}{;}{a}{}{,}

\usepackage[pagebackref=true]{hyperref}

\begin{document}

  \title{Foreground Mitigation and Power Spectrum Analysis for Tianlai Full-Sky 21\,cm Survey Observation
}
%   \subtitle{I. Place Your Subtitle Here}

   \volnopage{Vol.0 (20xx) No.0, 000--000}      %%preserved for Editor. DOn't remove!
   \setcounter{page}{1}          %%starting page, preserved for Editor. DOn't remove!

   \author{Yikai Deng %(周爱英) %% Put your Chinese name in "( )" if you like. Note to open line 11 "\usepackage[UTF8]{ctex}"
      \inst{1,2}
   \and Shifan Zuo
      \inst{1,2}
   \and Jixia Li
      \inst{1,2}
   \and Yougang Wang
      \inst{1,2,3}
   \and Xuelei Chen
      \inst{1,2,3}
   }

   \institute{State Key Laboratory of Radio Astronomy and Technology, National Astronomical Observatories, CAS, A20 Datun Road, Chaoyang District, Beijing, 100101, P. R. China; {\it sfzuo@bao.ac.cn}\\
        \and
             School of Astronomy and Space Science, University of Chinese Academy of Sciences, Beijing 100049, China\\
        \and
             Key Laboratory of Cosmology and Astrophysics (Liaoning) \& College of Sciences, Northeastern University, Shenyang 110819, China\\
\vs\no
   {\small Received 20xx month day; accepted 20xx month day}}

\abstract{We present a comprehensive analysis of the 21\,cm intensity mapping (IM) data from the Tianlai Cylinder Pathfinder Array (TCPA), focusing on multi-scale foreground mitigation and three-dimensional power spectrum estimation. Utilizing 20 days of drift-scan observations (714.4--781.7\,MHz, corresponding to H\,{\sc i} emission at redshift $z \approx 0.82$--$0.99$), we reconstruct high-fidelity sky maps by incorporating a high-precision, drone-measured primary beam model. This in-situ calibration significantly enhances reconstruction accuracy over previous analytical approximations. To address astrophysical foregrounds, which exceed the cosmological signal by approximately five orders of magnitude, we implement a robust multi-scale subtraction strategy—mPCA-UWTS—which combines an isotropic Undecimated Wavelet Transform on the Sphere (UWTS) with independent Principal Component Analysis (PCA) within each wavelet domain. We subsequently estimate the 3D power spectrum via Spherical Fourier--Bessel (SFB) decomposition, providing a mathematically rigorous treatment of wide-angle and line-of-sight curvature effects inherent in wide-field surveys. Our analysis demonstrates that the SFB framework effectively isolates systematic contaminants and recovers the clustering signal without the biases introduced by conventional flat-sky approximations. This work represents the first application of the SFB formalism to observational 21\,cm IM data, establishing it as a computationally efficient and scalable diagnostic tool for the next generation of wide-field 21\,cm surveys, including the Square Kilometre Array (SKA) and the full Tianlai array.
}

   % \authorrunning{Y.K. Deng, S.F. Zuo, J.X. Li., Y.G. Wang, X.L. Chen }            %author_head in even pages
   \authorrunning{Deng et al. 2026}
   \titlerunning{Tianlai Full-Sky 21\,cm Survey Data Analysis}  % title_head in odd pages

   \maketitle
%% The author head (on even pages) and the title head (on odd pages) will be
%% automatically extracted from \author{} and \title{}. Whenever the title is too long,
%% you will be asked to supply a shorter one by inserting either \authorrunning{} or
%% \titlerunning{} before \maketitle. Anyway, you can specify your own heads.
%%
%%
%% Note: In the following text body of your manuscript, please note several differences from
%%       other major journals:
%% (1) \subsection{Please Capitalize the First Letter of Each Notional Word in Subsection Title}
%% (2) Please Capitalize the First Letter of Each Notional Word in all tables' captions

%
%________________________________________________ sections below
%
\section{Introduction}           %% first-level sections will be auto-capitalized
\label{sec:intro}

The 21\,cm hyperfine transition of neutral hydrogen (H\,{\sc i}) has emerged as a preeminent cosmological probe, offering a unique window into the large-scale structure (LSS) of the Universe across a vast redshift range \citep{2021ApJ...918...56Z, 2024arXiv241108113P}. Unlike traditional galaxy surveys that detect individual luminous objects, 21\,cm intensity mapping (IM) measures the aggregate emission from multiple sources within a single voxel. This technique enables efficient mapping of the three-dimensional density field over enormous volumes \citep{2015aska.confE..19S, 2020PASP..132f2001L}, making it particularly well-suited for investigating dark energy, the Epoch of Reionization (EoR), and the growth of cosmic structure.

The primary challenge in 21\,cm IM is the presence of astrophysical foregrounds, which are typically four to five orders of magnitude more intense than the cosmological signal \citep{2015PhRvD..91h3514S, 2017MNRAS.464.4995M, 2018AJ....156...32E}. These foregrounds are dominated by diffuse Galactic synchrotron radiation, with additional contributions from free-free emission and extragalactic point sources \citep{2005ApJ...625..575S, 2021MNRAS.504..208C}. Fortunately, foregrounds exhibit spectral smoothness, which contrasts with the rapid frequency fluctuations of the 21\,cm signal. This spectral contrast underpins various mitigation strategies, including parametric methods like polynomial fitting \citep{2006ApJ...650..529W}, Gaussian Process Regression (GPR; \citealt{2018MNRAS.478.3640M}), and non-parametric blind signal separation techniques such as Principal Component Analysis (PCA; \citealt{2013MNRAS.434L..46S}), Independent Component Analysis (FastICA; \citealt{2012MNRAS.423.2518C}), and Generalized Morphological Component Analysis (GMCA) \citep{2020MNRAS.499..304C}. However, instrumental systematics—most notably frequency-dependent beam chromaticity—can ``leak'' foreground power into the cosmological window, significantly complicating the separation process \citep{2018MNRAS.476.3382A, 2019arXiv190912369C, 2021MNRAS.506.2041A, 2024RASTI...3..607G}.

Observational 21\,cm cosmology follows two complementary paths: global signal experiments and power spectrum interferometry. Global signal experiments, such as EDGES \citep{2018Natur.555...67B} and SARAS \citep{2022NatAs...6..607S}, utilize single-element radiometers to target sky-averaged absorption features from the ``Cosmic Dawn'' and EoR. Conversely, interferometric arrays like the Low-Frequency Array (LOFAR; \citealt{2021MNRAS.501....1G}), the Hydrogen Epoch of Reionization Array (HERA; \citealt{2023ApJ...945..124H}), and the Tianlai Cylinder Pathfinder Array \citep{2012IJMPS..12..256C} target the spatial fluctuations of 21\,cm emission. By mapping the 3D power spectrum, these instruments aim to resolve LSS and the Baryon Acoustic Oscillation (BAO) scale, providing critical constraints on the expansion history and growth of structure.

Modern 21\,cm interferometers often employ exceptionally wide fields of view (FoV) to achieve the survey volumes required for cosmological precision. Conventional analysis typically relies on $uv$-plane Fourier transforms, which are strictly valid only for narrow tracked fields where the flat-sky approximation holds. For wide-field drift-scan observations, the $m$-mode formalism \citep{2014ApJ...781...57S} uses spherical harmonic decomposition to naturally account for Earth's rotation and the all-sky nature of the data, providing a robust framework for sky map reconstruction from zenith-pointed arrays.

However, the wide-field nature of these surveys invalidates the traditional flat-sky approximation used in standard cosmological analyses. In the flat-sky limit, a single, fixed line-of-sight (LoS) is assumed for the entire survey volume—a simplification valid only for small patches of the sky. For wide-field and all-sky experiments, the LoS direction varies significantly across the survey area, leading to ``wide-angle effects'' that couple different physical scales and distort the recovered signal. Consequently, the conventional Cartesian Fourier power spectrum, $P(\mathbf{k})$, is susceptible to significant biases on large scales \citep{2024PhRvD.110f3524K}.

To overcome these geometric limitations, Spherical Fourier--Bessel (SFB) decomposition has emerged as a rigorous framework. Originally proposed in the 1990s \citep{1991MNRAS.249..678B, 1993cvf..conf..205L} and applied to galaxy surveys \citep[e.g.,][]{1995MNRAS.272..885F, 1995MNRAS.275..483H}, SFB provides an optimal basis for 3D fields on the sphere by combining spherical harmonics for angular distribution with spherical Bessel functions for the radial (redshift) dimension. While early adoption was limited by computational cost, recent optimizations \citep{2019arXiv190605866S} and efficient numerical codes \citep{2012A&A...540A..60L, 2021PhRvD.104l3548G} have made SFB analysis tractable. Studies have demonstrated the superiority of SFB over spherical harmonic tomography \citep{2014PhRvD..90f3515N, 2015A&A...578A..10L} and its robustness in measuring large-scale cosmological effects \citep{2025PhRvD.111b3522S}. Nevertheless, most SFB research has focused on galaxy surveys or simulations, with very few applications to observational 21\,cm IM data \citep{2016ApJ...833..242L}.

The Tianlai project is a pathfinder experiment designed to validate technologies for 21\,cm IM during the post-reionization epoch \citep{2012IJMPS..12..256C, 2015ApJ...798...40X, 2020SCPMA..6329862L, 2021MNRAS.506.3455W}. Located at the radio-quiet Hongliuxia site in Xinjiang, China, the experiment comprises two co-located interferometric arrays: a cylinder pathfinder and a dish pathfinder. The Tianlai Cylinder Pathfinder Array (TCPA) consists of three fixed parabolic cylindrical reflectors, each 15\,m (E-W) $\times$ 40\,m (N-S), operating in drift-scan mode. Currently targeting the 700--800\,MHz band ($z \approx 0.77$--$1.03$), the TCPA utilizes 96 dual-linear polarization feeds in a $31 + 32 + 33$ configuration. Feeds are uniformly spaced along the central 12.4\,m of the focal line to suppress grating lobes while maximizing sensitivity \citep{Zhangjiao2016}. The primary goal is to map the LSS and constrain cosmological parameters, particularly the dark energy equation of state.

In this paper, we analyze a 20-day observational dataset from the TCPA. We introduce an analysis pipeline that integrates a high-fidelity primary beam model derived from in-situ drone measurements with a multi-scale foreground mitigation strategy. We use an isotropic Undecimated Wavelet Transform on the Sphere (UWTS) combined with scale-dependent PCA for high-fidelity foreground removal. We then apply SFB decomposition to extract the 3D power spectrum, accounting for wide-angle and LoS curvature effects. This work demonstrates the viability of the SFB framework for extracting cosmological signals from real-world, wide-field interferometric datasets. The paper is organized as follows: Section \ref{sec:data} details data reduction, beam modeling, and foreground mitigation. Section \ref{sec:result} presents the SFB power spectrum results, followed by a discussion in Section \ref{sec:diss}. We conclude with a summary in Section \ref{sec:sum}.

%% Authors can give a citation as 'Michel et al. 1992'.
%% You may also use \cite, \citep and \citet for citation, and use Table~1 or Figure~1
%% and so forth. Using \ref and \label for cross-references of Tables/Figures
%% is a good way in adjusting/adding/removing text, tables or figures.

\section{Data Reduction and Analysis}
\label{sec:data}

\subsection{Observations and Data Reduction} 
\label{subsec:data}

We present an analysis of 20 days of drift-scan observations from the TCPA, spanning two epochs in 2018: January 21--February 3 (14 days) and March 22--27 (6 days). The processing pipeline starts with raw visibilities. Owing to storage and throughput limitations during commissioning, the analyzed bandwidth is restricted to 712.9--783.1\,MHz, divided into 576 channels (0.122\,MHz width). Raw data underwent automated RFI flagging and calibration using Cygnus A, supplemented by an on-site artificial calibration noise source (CNS) for phase stabilization. Nightly mean subtraction and coupling corrections were applied to the calibrated visibilities. During map-making, bright celestial sources—Cassiopeia A, Cygnus A, and Centaurus A—were modeled and subtracted. Additionally, approximately 30 minutes of data around the Sun's daily transit were excised to mitigate sidelobe contamination. The $m$-mode map-making framework was presented in \citet{2014ApJ...781...57S,Zhangjiao2016a}, the detailed pipeline and calibration strategy and are documented in \citet{2021A&C....3400439Z}, \citet{2019AJ....157...34Z} respectively, and the error in the calibration were also analyzed by simulation in \citet{YuKaifeng2023}.
We reconstruct independent sky maps for the XX and YY polarizations to account for their distinct primary beam responses.

Previous studies utilized the analytical beam model in \texttt{tlpipe} \citep{2021A&C....3400439Z}, which assumes a separable profile in the east--west (EW) and north--south (NS) directions \citep{2014ApJ...781...57S}. However, simulations \citep{2022RAA....22f5020S} and in-situ drone measurements \citep{2025arXiv250801413L} demonstrate that the actual beam response deviates significantly from this idealized form. To enhance reconstruction accuracy, we utilize a high-precision beam profile for the A26 feed, derived from drone-based far-field measurements \citep{2025arXiv250801413L}. This model covers the 693.24--801.64\,MHz range with 0.98\,MHz resolution.

Due to drone operational constraints, measurements were limited to zenith angle ranges of $|\theta_{\text{NS}}| < 70^{\circ}$ and $|\theta_{\text{EW}}| < 3^{\circ}$. To construct a comprehensive all-sky model, we extend these to $\pm 90^{\circ}$ in both directions across all frequencies. In the measured EW domain, we use cubic spline interpolation for fine-scale structures. For unmeasured regions, a piecewise Gaussian model accounts for beam asymmetry; the two sides of the EW profile are fitted independently, with scaling factors applied to the Gaussian tails for continuity. For the NS direction, the sparsity and irregularity of the data necessitate a single Gaussian fit over the $|\theta_{\text{NS}}| < 70^{\circ}$ range.

\begin{figure*}
\centering
\begin{minipage}{0.49\linewidth}
\centering\includegraphics[width=\textwidth]{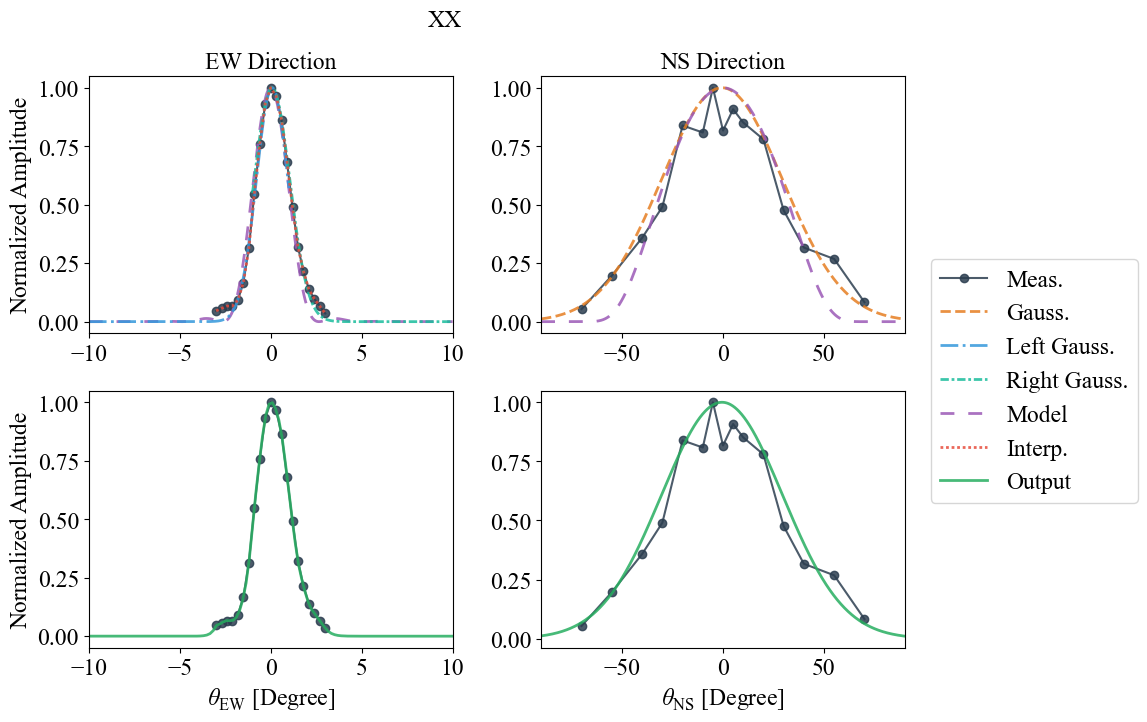}
\end{minipage}
\hfill
\begin{minipage}{0.49\linewidth}
\centering\includegraphics[width=\textwidth]{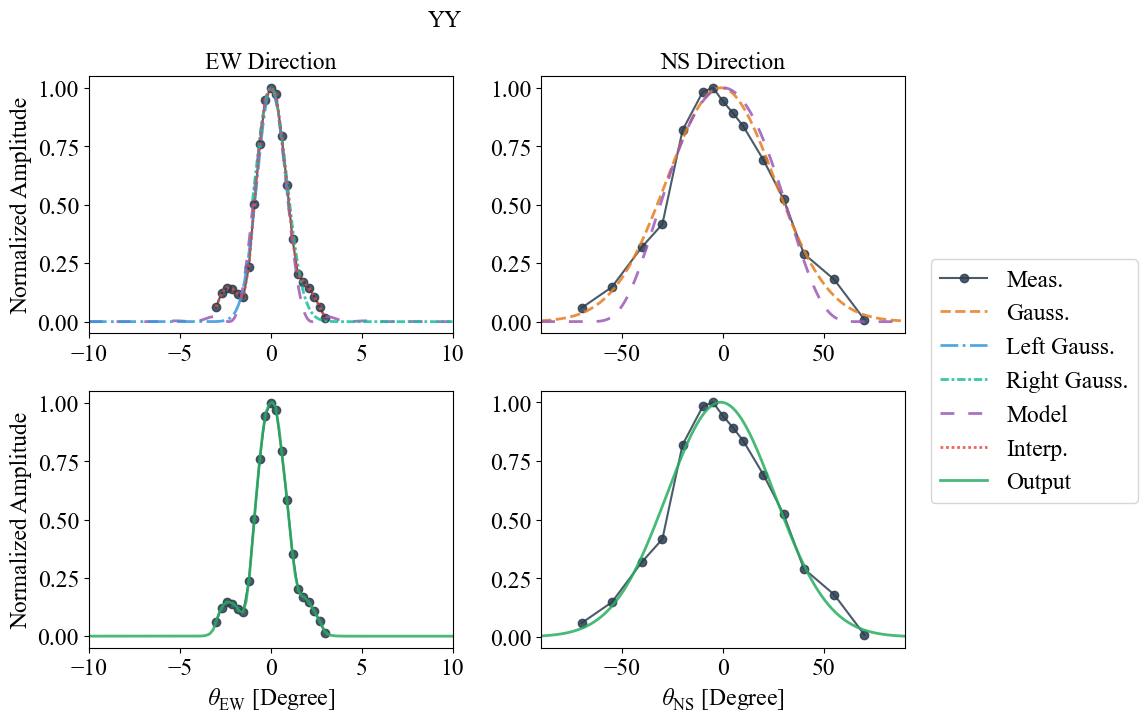}
\end{minipage}
\caption{One-dimensional primary beam response for the A26 feed as a function of zenith angle at a representative frequency (748 MHz) for the XX (left panels) and YY (right panels) polarizations. Drone-based measurements (dot markers) are compared with three modeling approaches: a Gaussian fit, the default analytical model from \texttt{tlpipe}, and cubic spline interpolation. To account for beam asymmetry, the two sides of the EW profile are fitted independently, as legend with Left Guass. and Right Gauss., respectively. The upper panels illustrate the fitting performance within the measured domain, while the lower panels show the extended all-sky beam profiles in the interested angle range for clarity.
\label{fig:vs1d}}
\end{figure*}

\begin{figure*}
\centering
\begin{minipage}{0.24\linewidth}
\centering\includegraphics[width=\textwidth]{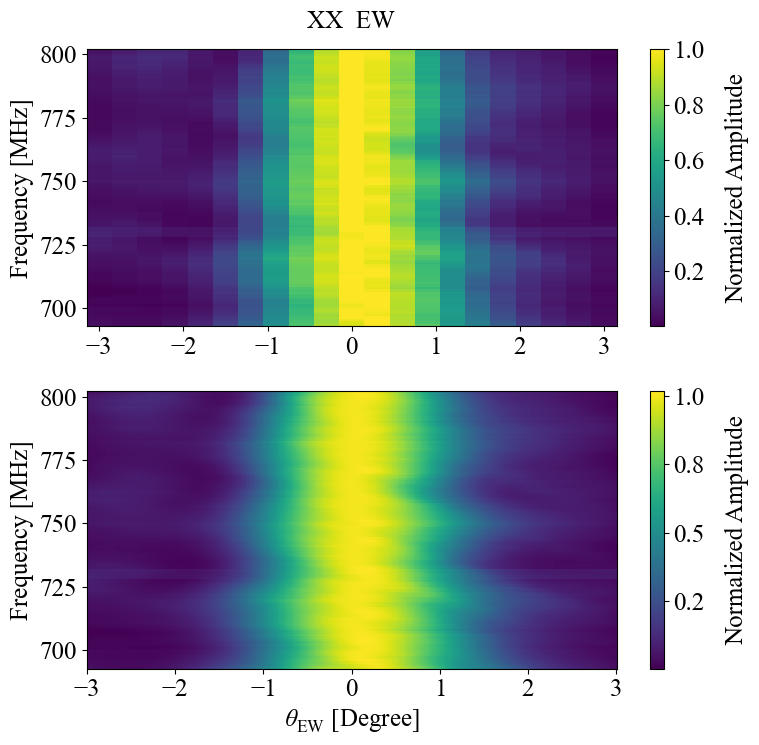}
\end{minipage}
\hfill
\begin{minipage}{0.24\linewidth}
\centering\includegraphics[width=\textwidth]{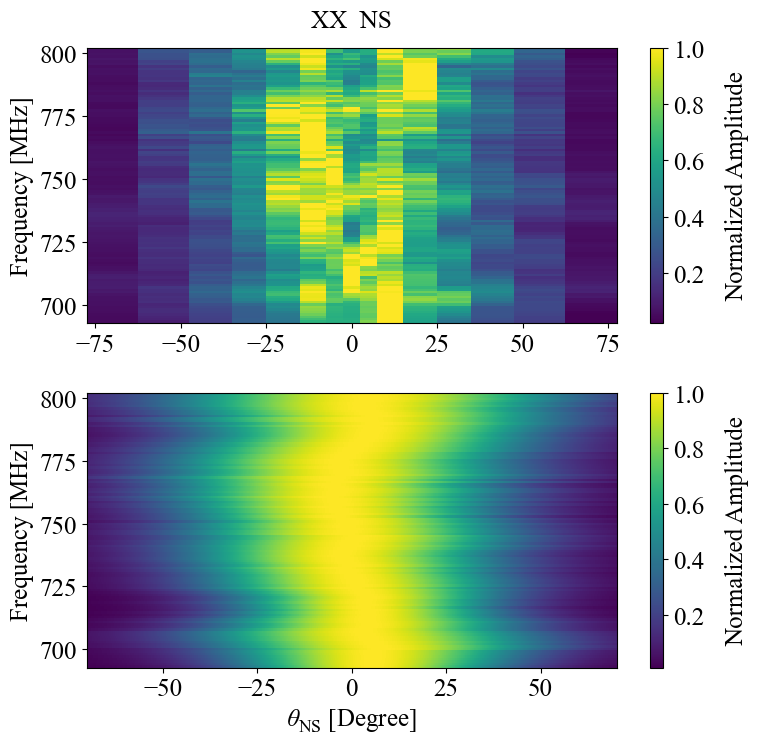}
\end{minipage}
\hfill
\begin{minipage}{0.24\linewidth}
\centering\includegraphics[width=\textwidth]{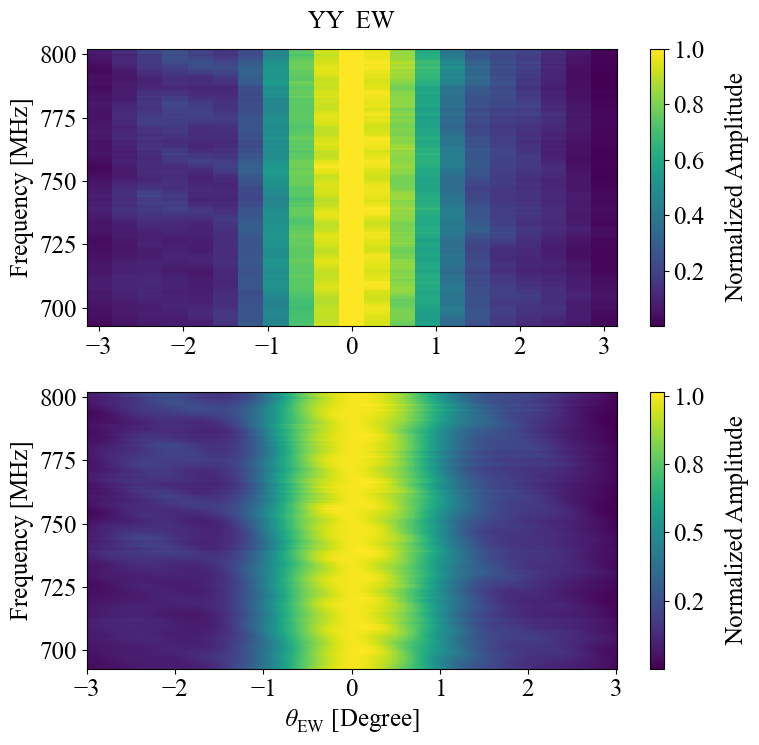}
\end{minipage}
\hfill
\begin{minipage}{0.24\linewidth}
\centering\includegraphics[width=\textwidth]{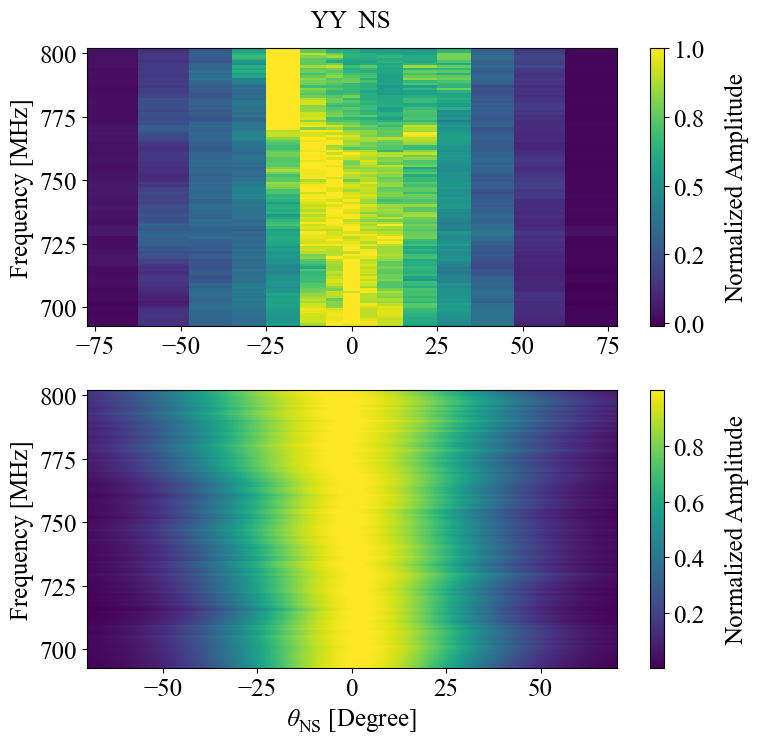}
\end{minipage}
\caption{Frequency-dependent evolution of the drone-measured beam response. The top row displays the raw far-field measurements across the observed band. The bottom row shows the resulting models: cubic spline interpolation for the EW direction and Gaussian fits for the NS direction.
\label{fig:fit2d}}
\end{figure*}

\begin{figure*}
\centering
\begin{minipage}{0.24\linewidth}
\centering\includegraphics[width=\textwidth]{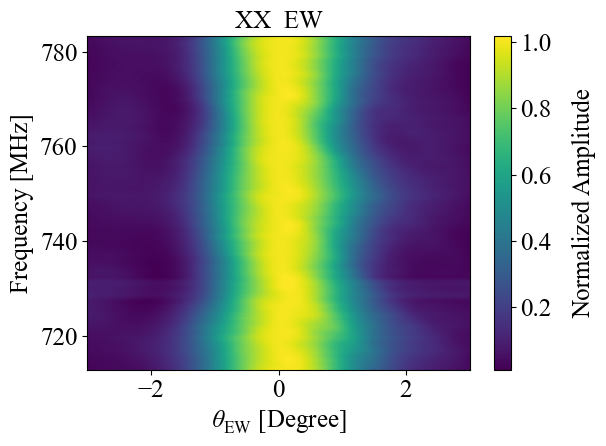}
\end{minipage}
\hfill
\begin{minipage}{0.24\linewidth}
\centering\includegraphics[width=\textwidth]{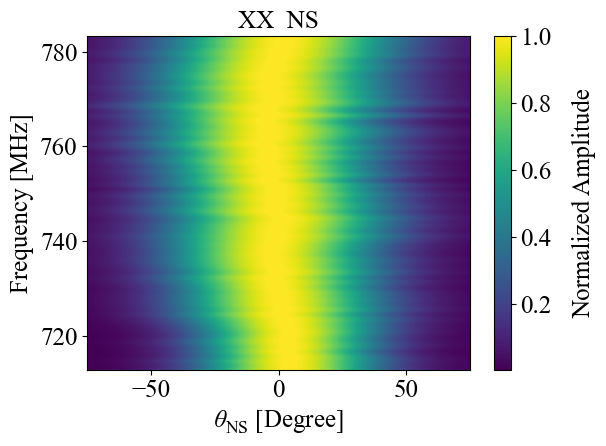}
\end{minipage}
\hfill
\begin{minipage}{0.24\linewidth}
\centering\includegraphics[width=\textwidth]{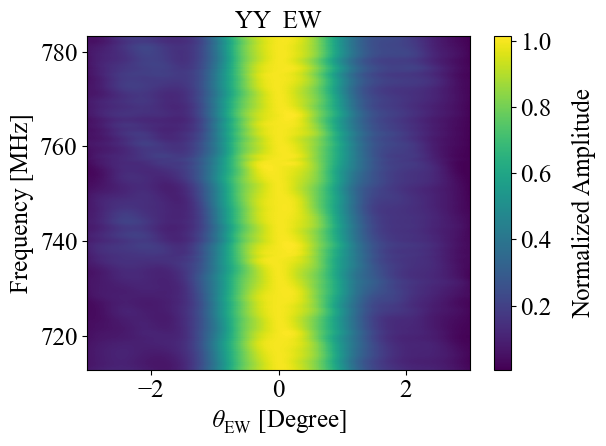}
\end{minipage}
\hfill
\begin{minipage}{0.24\linewidth}
\centering\includegraphics[width=\textwidth]{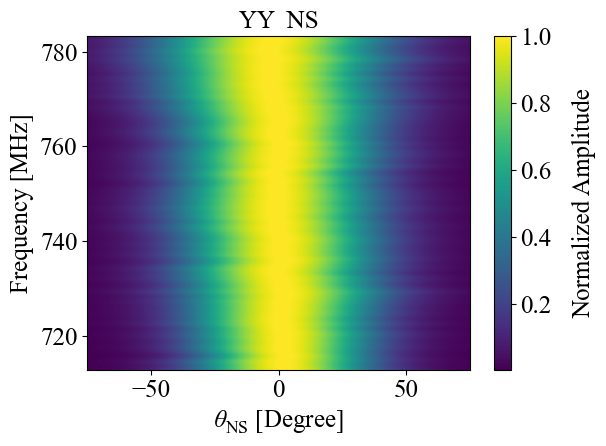}
\end{minipage}
\caption{Interpolation of the sparse drone-measured beam profiles to the full frequency resolution (576 channels) of the visibility data. This ensures a consistent beam model for all frequency channels used in the map-making process.
\label{fig:inp2d}}
\end{figure*}

\begin{figure*}
\centering
\begin{minipage}{0.32\linewidth}
\centering\includegraphics[width=\textwidth]{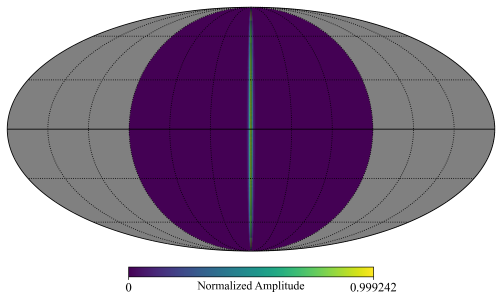}
\end{minipage}
\hfill
\begin{minipage}{0.32\linewidth}
\centering\includegraphics[width=\textwidth]{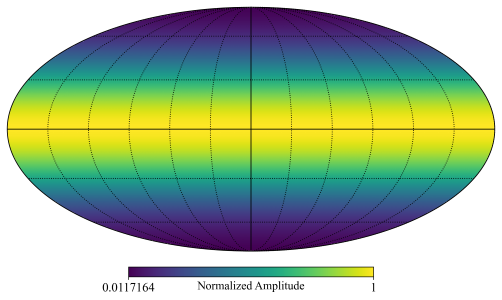}
\end{minipage}
\hfill
\begin{minipage}{0.32\linewidth}
\centering\includegraphics[width=\textwidth]{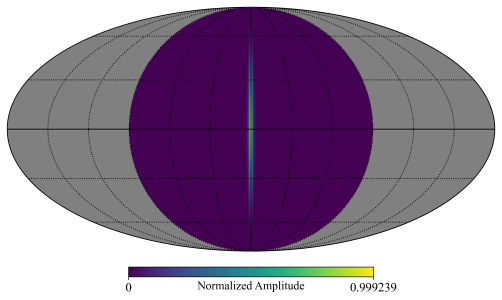}
\end{minipage}
\vfill
\begin{minipage}{0.32\linewidth}
\centering\includegraphics[width=\textwidth]{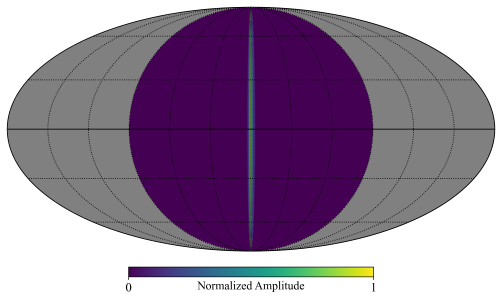}
\end{minipage}
\hfill
\begin{minipage}{0.32\linewidth}
\centering\includegraphics[width=\textwidth]{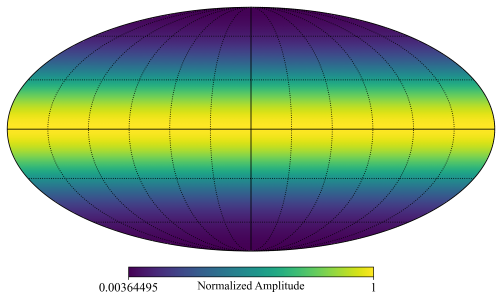}
\end{minipage}
\hfill
\begin{minipage}{0.32\linewidth}
\centering\includegraphics[width=\textwidth]{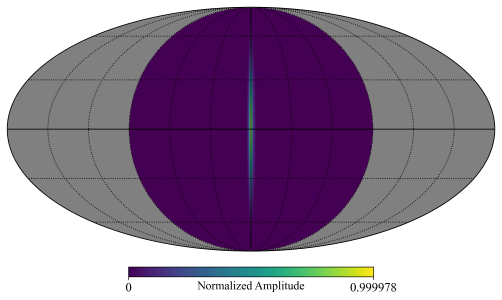}
\end{minipage}
\caption{Two-dimensional primary beam model projected onto the celestial sphere. The EW (left) and NS (middle) profiles are displayed in HEALPix Mollview projection at the central frequency (748\,MHz). The final 2D model (right), constructed assuming separability, is used for wide-field map reconstruction. Top and bottom rows correspond to the XX and YY polarizations, respectively.
\label{fig:proj}}
\end{figure*}

Figure~\ref{fig:vs1d} shows the beam profile along the East-West and North-South directions as measured by the drone and fitting models for a representative frequency.  The full frequency-dependent profile is shown in Figure~\ref{fig:fit2d}. To align these with the 576 frequency channels of our visibility data, we implement a multi-stage interpolation. First, we shift the peaks of the fitted profiles to a common zenith angle to avoid interpolation artifacts from minor frequency-dependent offsets in the raw data. These offsets are then re-applied to the interpolated profiles (Figure~\ref{fig:inp2d}). The final EW and NS profiles are projected onto the celestial sphere using the HEALPix \citep{2005ApJ...622..759G} Mollview projection (Figure~\ref{fig:proj}). Assuming beam separability, our 2D primary beam model is constructed as the product of these orthogonal 1D profiles.

\begin{figure*}
\centering
\begin{minipage}{0.48\linewidth}
\centering\includegraphics[width=\textwidth]{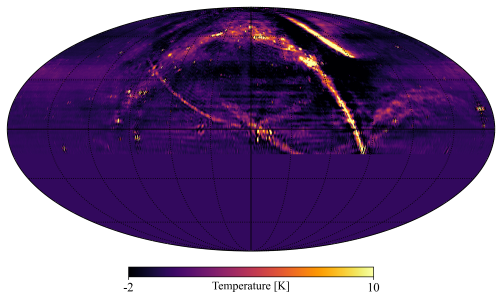}
\end{minipage}
\hfill
\begin{minipage}{0.48\linewidth}
\centering\includegraphics[width=\textwidth]{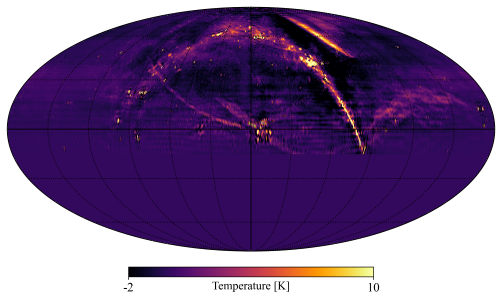}
\end{minipage}
\vfill
\begin{minipage}{0.48\linewidth}
\centering\includegraphics[width=\textwidth]{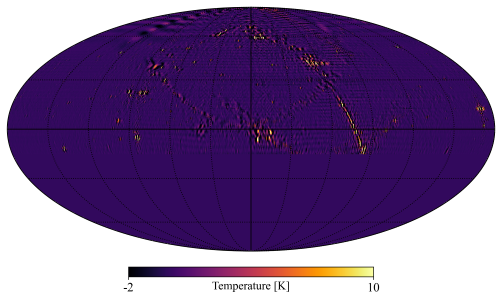}
\end{minipage}
\hfill
\begin{minipage}{0.48\linewidth}
\centering\includegraphics[width=\textwidth]{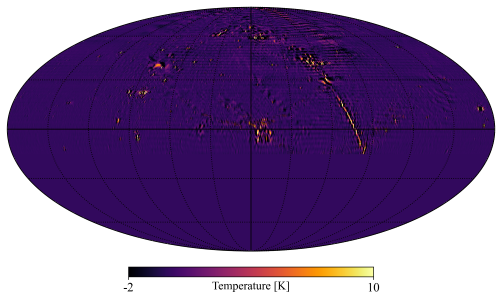}
\end{minipage}
\caption{Reconstructed sky maps for the XX (left) and YY (right) polarizations at 748\,MHz. The top row shows maps using the all-baseline set, while the bottom row utilizes only the cross-cylinder baseline subset. The maps are dominated by Galactic diffuse emission and systematic artifacts (e.g., solar sidelobes and Galactic ghosts). The temperature scale is restricted to $[-2, 10]$\,K to highlight faint structures.
\label{fig:map}}
\end{figure*}

Figure~\ref{fig:map} (top panels) shows the reconstructed sky maps for the XX (left) and YY (right) polarizations at 748\,MHz. The maps are dominated by Galactic diffuse emission and bright extragalactic point sources. We identify a prominent ``ghost'' artifact—an aliased image of the Galactic plane at the upper center-right of the figure, resulting from incomplete angular sampling. Additionally, two parabolic arcs from solar sidelobes are visible near the center of the figure, caused by the Sun's shift between the two 2018 epochs. Despite excising 30 minutes of peak transit data, residual solar power remains detectable. Failure to remove the peak solar signal would cause severe contamination across the map via the instrument's Point Spread Function (PSF).

The Maps reconstructed with this drone-measured beam exhibit improved fidelity over those using the default analytic model. Given the array's geographic latitude ($44^{\circ}09'$N), sensitivity drops precipitously south of $\delta \approx -15^{\circ}$. Consequently, maps become noise-dominated at colatitudes $\theta > 105^{\circ}$. We restrict subsequent cosmological analyses to the colatitude range $[0^{\circ}, 105^{\circ}]$. Additionally, frequency channels at both ends of the band are discarded to mitigate edge effects and band-pass instabilities. The final science-ready data cube comprises $N_\nu = 552$ channels spanning 714.42--781.68\,MHz with $N_{\text{side}}=512$ resolution. Note that in this figure the $a_{00}$ spherical harmonic component (global mean temperature $\bar{T}$) is filtered out to suppress RFI and common-mode pickup, resulting in zero-mean maps.

We also reconstruct maps using only cross-cylinder baselines (Figure~\ref{fig:map}, bottom panels). These longer baselines ($\ge 15.0$\,m) are less susceptible to crosstalk. While the Galactic plane remains visible, its large-scale structure is attenuated by the lack of short baselines (the ``missing-zero'' problem). Notably, systematic artifacts like the Galactic ghost and solar sidelobes are markedly suppressed in these maps, though not entirely eliminated.

For comparison, we have also generated a simulated 21\,cm signal using \texttt{cora} \citep{2025zndo..17438047S} within the same frequency band, applying a declination cut at $\delta = -15^{\circ}$ to match the Tianlai observational footprint.

\subsection{Foreground Subtraction} 
\label{subsec:fore}

The reconstructed sky maps are dominated by intense diffuse foreground emission, which must be accurately characterized and removed to extract the faint cosmological 21\,cm signal. 

Conventional foreground mitigation techniques, such as polynomial fitting \citep{2006ApJ...650..529W}, Principal Component Analysis \citep[PCA;][]{2013MNRAS.434L..46S}, Fast Independent Component Analysis \citep[FastICA;][]{2012MNRAS.423.2518C}, and Gaussian Process Regression \citep[GPR;][]{2018MNRAS.478.3640M, 2022MNRAS.510.5872S}, primarily exploit the spectral smoothness of astrophysical foregrounds \citep{2021MNRAS.504..208C, 2019arXiv190912369C}. However, incorporating spatial (angular) information is essential, as instrumental systematics and astrophysical contaminants often exhibit distinct morphological characteristics across different spatial scales. For instance, while large-scale features are dominated by synchrotron and free-free emission, small-scale fluctuations are typically more sensitive to instrumental noise and localized systematics.

Drawing upon the multi-scale PCA (mPCA) algorithm proposed by the MeerKLASS Collaboration \citep{2025A&A...703A.222C}, we implement a tailored approach for wide-field spherical maps. Within the mPCA framework, each frequency map is first decomposed into multiple wavelet-filtered components representing distinct spatial scales via the isotropic undecimated wavelet transform (IUWT, or starlet transform) \citep{2007ITIP...16..297S}. A temperature map $X(\nu, p)$ at frequency $\nu$ and pixel $p$ is expressed as the sum of a coarse-scale component $X_{\text{L}}(\nu, p)$ and a set of wavelet scales $W_{j}(\nu, p)$:
\begin{equation} \label{eq:Xvp}
  X(\nu, p) = X_{\text{L}}(\nu, p) + \sum_{j=1}^{j_{\text{max}}} W_{j}(\nu, p).
\end{equation}
The wavelet coefficients $W_{j}$ capture features at dyadic scales $2^{j}$, while $X_{\text{L}}$ represents the remaining large-scale information. 
In contrast to the original MeerKLASS implementation, which utilized a single wavelet scale for small-area flat-sky maps, our analysis adopts a multi-scale decomposition to accommodate the wide-field geometry and complex systematic environment of the TCPA dataset.

For spherical datasets, we utilize the isotropic Undecimated Wavelet Transform on the Sphere (UWTS) \citep{2006A&A...446.1191S}. The UWTS retains the key advantages of the flat-sky starlet transform—including exact reconstruction, isotropy, and compact support—while operating directly in the spherical harmonic domain. The decomposition is performed using a scaling function $\phi_{l_{c}}(\theta, \varphi)$ characterized by azimuthal symmetry and a cut-off multipole $l_{c}$. The scaling function can be choose in harmonic space as
\begin{equation} 
  \hat{\phi}_{l_c}(l) =\frac{3}{2}B_3(\frac{2l}{l_c}),
\end{equation}
where $B_3(x)=\frac{1}{12}(|x-2|^3-4|x-1|^3+6|x|^3-4|x+1|^3+|x+2|^3)$ is a B-spline function of order 3, which is very close to a Gaussian function and converges rapidly to 0. A sequence of smoother approximations of a function $f$ on a dyadic resolution scale can be obtained using the scaling function $\phi_{l_{c}}$ as
\begin{equation} 
  c_{j} = \phi_{2^{-j}l_c} * f,
\end{equation}
where $*$ stands for convolution.
The wavelet coefficients can be defined as the difference between two consecutive resolutions
\begin{equation} 
  w_{j+1} = c_j - c_{j+1}.
\end{equation}

In this work, we set $l_{c} = 1024$, matching the maximum angular resolution accessible to the TCPA ($\ell \sim 750$). This yields $j_{\text{max}} = 10$, resulting in 10 wavelet scales $w_1$ to $w_{10}$ and one coarse component $c_{10}$ for each sky map. We then apply PCA to each scale independently. By organizing the data at each scale into a 2D matrix $\mathbf{T}$ of dimension $N_\nu \times N_{\text{pix}}$, we perform Singular Value Decomposition (SVD):
\begin{equation}
\mathbf{T} = \mathbf{U} \mathbf{\Sigma} \mathbf{V}^{\text{T}},
\end{equation}
where the columns of $\mathbf{U}$ represent the principal spectral modes and $\mathbf{\Sigma}$ denotes the singular values $s_i$. The eigenvalues of the frequency covariance matrix $\mathbf{C} = \frac{\mathbf{T} \mathbf{T}^{\text{T}}}{N_{\text{pix}}-1}$ are given by $\lambda_i = s_i^2$. 

The foreground-cleaned residual for each scale is obtained by subtracting the $N_{\text{FG}}$ dominant modes:
\begin{equation}
\mathbf{T}_{\text{res}} = \mathbf{T} - \sum_{i=1}^{N_{\text{FG}}} s_i \mathbf{u}_i \mathbf{v}_i^{\text{T}},
\end{equation}
where $N_{\text{FG}}$ is determined independently for each scale to optimize the balance between foreground suppression and signal preservation. The final cleaned map is reconstructed by summing the residuals across all scales:
\begin{equation}
T_{\text{clean}}(\hat{\bm{r}}, \nu) = c_{j_{\text{max}}, \text{res}}(\hat{\bm{r}}, \nu) + \sum_{j=1}^{j_{\text{max}}} w_{j, \text{res}}(\hat{\bm{r}}, \nu).
\end{equation}
The exact reconstruction property of the UWTS ensures that the decomposition-recomposition process introduces no numerical artifacts.
% ; we have verified that the numerical reconstruction error is $< 10^{-5}$\,K. 

A key advantage of mPCA over conventional PCA is the ability to adapt the number of removed components to the specific noise and foreground properties of each spatial scale. However, this flexibility necessitates a robust criterion for selecting $N_{\text{FG}}$. In this study, we utilize the {\it fractional variance explained} as the primary metric for determining the optimal $N_{\text{FG}}$ at each scale.

\begin{figure*}
\centering
\begin{minipage}{0.48\linewidth}
\centering\includegraphics[width=\textwidth]{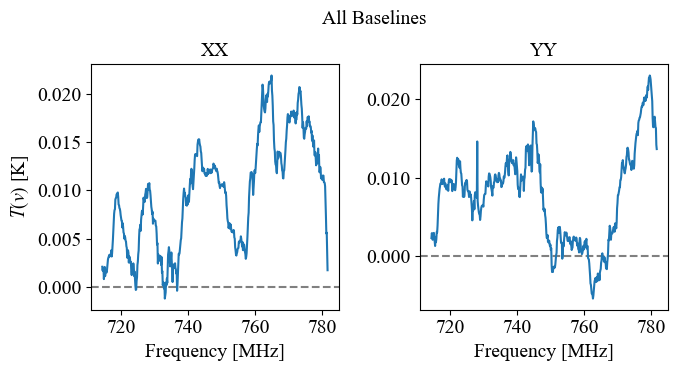}
\end{minipage}
\hfill
\begin{minipage}{0.48\linewidth}
\centering\includegraphics[width=\textwidth]{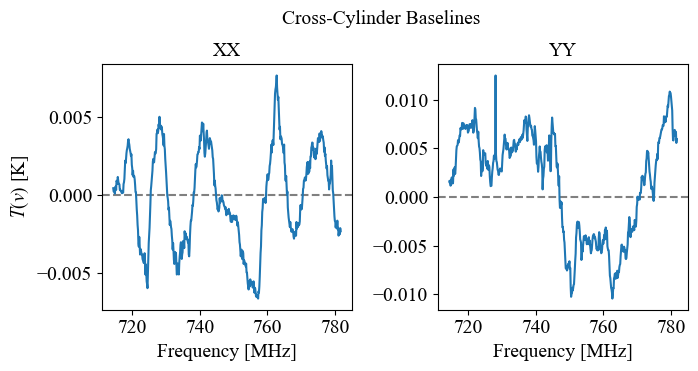}
\end{minipage}
\caption{Mean frequency spectra for the reconstructed XX and YY polarization sky maps. The left panel shows the results using the all-baseline set, while the right panel corresponds to the cross-cylinder baseline subset.
\label{fig:fspec}}
\end{figure*}

Figure \ref{fig:fspec} illustrates the mean frequency spectra for the XX and YY polarizations. The YY polarization exhibits a markedly smoother spectral profile compared to XX, rendering it more amenable to foreground mitigation. Given that the cosmological 21\,cm signal is expected to be unpolarized, we focus our subsequent power spectrum analysis on the YY polarization maps.

\subsection{Spherically Averaged Power Spectrum Estimation} 
\label{subsec:pk}

The spherically averaged power spectrum, $P(k)$, quantifies the statistical variance of 21\,cm brightness temperature fluctuations across different spatial scales. It serves as a fundamental observable for extracting cosmological and astrophysical constraints from intensity mapping data. In the flat-sky approximation—commonly applied to narrow-field observations—the power spectrum is typically decomposed into components parallel ($k_\parallel$) and perpendicular ($k_\perp$) to a fixed line-of-sight (LoS) \citep{2016ApJ...833..242L}. Assuming statistical isotropy, the power spectrum depends solely on the magnitude of the 3D wavevector $\bm{k}$, where $k = \sqrt{k_\perp^2 + k_\parallel^2}$.

Standard Fourier-based estimators, which involve a 3D Fast Fourier Transform (FFT) of a Cartesian image cube, are not directly applicable to our data given the wide-field coverage and the spherical shell geometry spanning the redshift range $z \approx 0.82$--$0.99$. Instead, we adopt the methodology proposed by \citet{2007MNRAS.378..119D} and \citet{2022MNRAS.516.2851P} to estimate the spherically averaged $P(k)$. We first compute the multi-frequency angular power spectrum (MAPS), $C_\ell(\nu_a, \nu_b)$, defined as:
\begin{equation}
C_\ell(\nu_a, \nu_b) = \langle a_{\ell m}(\nu_a) a_{\ell m}^*(\nu_b) \rangle.
\end{equation}
Assuming statistical homogeneity over the narrow relative bandwidth ($\Delta \nu \ll \nu_c$), the MAPS depends primarily on the frequency separation $\Delta \nu = |\nu_b - \nu_a|$. We average the $C_\ell$ values for frequency pairs with consistent $\Delta \nu$ to obtain $C_\ell(\Delta \nu)$. The cylindrically binned power spectrum $P(k_\perp, k_\parallel)$ is then derived via a Fourier transform of $C_\ell(\Delta \nu)$ along the frequency axis:
\begin{equation}
P(k_\perp, k_\parallel) = r^2 r' \int d(\Delta \nu) e^{-i k_\parallel r' \Delta \nu} C_\ell(\Delta \nu),
\end{equation}
where $k_\perp = \ell / r$. At the reference frequency $\nu_c = 748$\,MHz, the comoving distance $r$ and its derivative $r' = dr/d\nu$ are $2151.15\,h^{-1}$\,Mpc and $4.66\,h^{-1}$\,Mpc\,MHz$^{-1}$, respectively, calculated using the WMAP9 cosmology \citep{2013ApJS..208...19H}. Finally, the 1D power spectrum $P(k)$ is obtained by spherically averaging the 2D $P(k_\perp, k_\parallel)$ into 20 $k$-bins with a uniform logarithmic width of $\Delta \log_{10} k \approx 0.1$, spanning the range $k \in [0.002, 0.2]\,h\,\text{Mpc}^{-1}$.

\subsection{Spherical Fourier--Bessel Power Spectrum} 
\label{subsec:clnn}

In observational cosmology, spherical coordinates—centered on the observer—provide the most natural basis for analyzing large-scale structure (LSS) fields. In this geometry, the eigenfunctions of the Laplacian operator are the product of spherical Bessel functions and spherical harmonics, $j_{\ell}(kr)Y_{\ell m}(\hat{\bm{r}})$, with corresponding eigenvalues $-k^{2}$. This basis framework intrinsically accounts for sky curvature and the expanding survey volume, avoiding the wide-angle biases inherent in flat-sky approximations. For a statistically homogeneous 3D field $\delta(\bm{r})$ in a flat universe, the spherical Fourier--Bessel (SFB) decomposition \citep{1995MNRAS.272..885F, 2003MNRAS.343.1327H, 2005PhRvD..72b3516C} is defined as:
\begin{equation} \label{eq:dr}
  \delta(\bm{r}) = \int dk \sum_{\ell m} \left[ \sqrt{\frac{2}{\pi}} k j_{\ell}(kr) Y_{\ell m}(\hat{\bm{r}}) \right] \tilde{\delta}_{\ell m}(k),
\end{equation}
with the inverse relation:
\begin{equation} \label{eq:dlm}
  \tilde{\delta}_{\ell m}(k) = \int d^{3}r \left[ \sqrt{\frac{2}{\pi}} k j_{\ell}(kr) Y^{*}_{\ell m}(\hat{\bm{r}}) \right] \delta(\bm{r}).
\end{equation}
The SFB power spectrum $C_{\ell}(k)$ is defined via the covariance:
\begin{align} \label{eq:clk}
  \langle \tilde{\delta}_{\ell m}(k) \tilde{\delta}^{*}_{\ell' m'}(k')\rangle &= C_{\ell}(k,k')\delta_{\ell\ell'}\delta_{mm'}\\ \notag
  &= C_{\ell}(k) \delta^{D}(k - k')\delta_{\ell\ell'}\delta_{mm'}.
\end{align}

\begin{figure}
\centering
\includegraphics[width=0.48\textwidth]{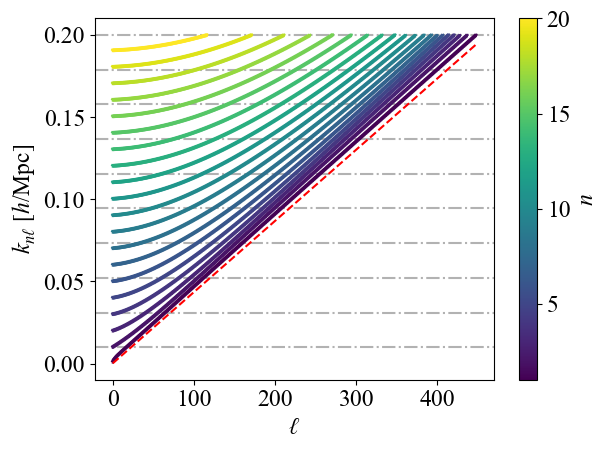}
\caption{Distribution of the discrete SFB modes $k_{n\ell}$ as a function of angular multipole $\ell$ and radial index $n$, calculated for the redshift range $z \in [0.8, 1.0]$. The dashed red line indicates the boundary formed by the $n=1$ modes, representing the largest radial scales accessible to the survey. This boundary is well-approximated by the geometric relation $k \approx (\ell + 1/2) / r_{\text{max}}$.
\label{fig:knl}}
\end{figure}

Practical surveys are confined to finite volumes, in which case the radial modes have a discrete set of values $k_{n\ell}$. For a survey volume defined by a spherical shell with boundaries $r_{\text{min}}$ and $r_{\text{max}}$, the SFB decomposition of the temperature field $T(\bm{r})$ is:
\begin{equation}
  T_{n\ell m} = \int d^2\hat{\bm{r}} \, Y_{\ell m}^*(\hat{\bm{r}}) \int_{r_{\text{min}}}^{r_{\text{max}}} dr \, r^2 g_{n\ell}(r) T(\bm{r}),
\end{equation}
where the radial basis functions $g_{n\ell}(r)$ are linear combinations of spherical Bessel functions of the first ($j_\ell$) and second ($y_\ell$) kind:
\begin{equation}
  g_{n\ell}(r) = c_{n\ell} j_\ell(k_{n\ell} r) + d_{n\ell} y_\ell(k_{n\ell} r).
\end{equation}
These functions, which are the general solution of the radial part of Laplace's equation, i.e., the spherical Bessel equation, satisfy the orthonormality relation:
\begin{equation} \label{eq:orth}
  \int_{r_{\text{min}}}^{r_{\text{max}}} dr \, r^2 g_{n\ell}(r)g_{n'\ell}(r) = \delta_{nn'}.
\end{equation}
The coefficients $c_{n\ell}$ and $d_{n\ell}$ and the discrete mode $k_{n\ell}$ are determined by the orthonormality and the boundary conditions at $r_{\text{min}}$ and $r_{\text{max}}$. 
In this analysis, we adopt the potential boundary condition \citep{1995MNRAS.272..885F}, which ensures the field represented by the SFB decomposition is continuous and smooth at the boundary. Other boundary conditions, for example, the velocity boundary condition \citep{1995MNRAS.272..885F}, which sets the derivative of the density field, that is the velocity, to vanish on the boundaries, can also be used. The resulting discrete radial eigenvalues $k_{n\ell}$ are coupled to both the radial index $n$ and the angular multipole $\ell$, as illustrated in Figure~\ref{fig:knl}. For the TCPA frequency band, the comoving boundaries are $r_{\text{min}} = 1997.83\,h^{-1}$\,Mpc and $r_{\text{max}} = 2310.81\,h^{-1}$\,Mpc.

The discrete modes $k_{n\ell}$ represent the magnitude of the wavevector $\bm{k}$, analogous to $k$ in the Cartesian Fourier basis. The angular transform is performed via the HEALPix framework, while the radial integration is implemented using the trapezoidal rule. The observed (pseudo) SFB power spectrum, $\hat{C}_{\ell nn'}$, is then estimated as:
\begin{equation}
\hat{C}_{\ell nn'} = \frac{1}{2\ell + 1} \sum_{m} T_{n\ell m} T_{n'\ell m}^*.
\end{equation}
Assuming statistical isotropy, we average over the $m$-modes and focus on the diagonal components ($n = n'$). The pseudo-power spectrum extends the standard angular power spectrum $C_\ell$ to three dimensions.

To mitigate the effects of an incomplete survey volume, window function deconvolution is required to recover an unbiased estimate of the underlying power spectrum. Under the assumption of a separable observational window function $W(\bm{r}) = \phi(r) M(\hat{\bm{r}})$, where $\phi(r)$ and $M(\hat{\bm{r}})$ represent the radial and angular selection functions respectively, the observed SFB power spectrum $C_{\ell nn'}^{\text{obs}}$ is related to the true spectrum $C_{LN N'}$ via a mixing matrix $\mathcal{M}$ \citep{2021PhRvD.104l3548G}:
\begin{equation} \label{eq:wf}
C_{\ell nn'}^{\text{obs}} = \sum_{LN N'} \mathcal{M}^{LN N'}_{\ell nn'} C_{LN N'},
\end{equation}
where the mixing matrix is defined as:
\begin{equation} \label{eq:M}
\begin{split}
\mathcal{M}^{LN N'}_{\ell nn'} = \, & \frac{2L+1}{4\pi} \sum_{L_1} 
\begin{pmatrix} 
\ell & L & L_1 \\ 
0 & 0 & 0 
\end{pmatrix}^2 \sum_{M_1} |W_{L_1 M_1}|^2 \\ 
                                 & \times \int dr \, r^2 g_{n\ell}(r) g_{NL}(r) \phi(r) \\
  & \times \int dr' \, r'^2 g_{n'\ell}(r') g_{N'L}(r') \phi(r').
\end{split}
\end{equation}
Here, $W_{L_1 M_1}$ are the spherical harmonic coefficients of the angular mask $M(\hat{\bm{r}})$, and the term in parentheses is the Wigner $3j$ symbol. For our survey, we set $\phi(r) = 1$ and $M(\hat{\bm{r}}) = 1$ within the colatitude range $\theta < 105^{\circ}$, and zero otherwise. Deconvolution aims to retrieve the true spectrum $C_{LN N'}$ from the observed $C_{\ell nn'}^{\text{obs}}$, though this inversion is often numerically ill-conditioned.

To suppress the variance of poorly constrained modes and compress the data, we group the SFB pseudo-power spectrum into bandpowers. The binning is performed over both $\ell$ and $n$ modes to obtain the bandpower-binned SFB power spectrum:
\begin{equation} \label{eq:bb}
B_{LN N'}^{\text{obs}} = \sum_{\ell nn'} \widetilde{w}_{LN N'}^{\ell nn'} C_{\ell nn'}^{\text{obs}},
\end{equation}
where the weights $\widetilde{w}$ average neighboring modes $(\ell n n') \sim (L N N')$ and are normalized such that $\sum_{\ell nn'} \widetilde{w}_{LN N'}^{\ell nn'} = 1$. In matrix notation, the relationship between the observed bandpowers $B^{\text{obs}}$ and the underlying spectrum $C$ is:
\begin{equation}
B^{\text{obs}} = \widetilde{w} \mathcal{M} C = \mathcal{N} B,
\end{equation}
where $\mathcal{N} = \widetilde{w} \mathcal{M} v$ is the bandpower mixing matrix and $v$ is the Moore--Penrose inverse of $\widetilde{w}$. The bin widths $\Delta \ell$ and $\Delta n$ are estimated based on the survey sky fraction $f_{\text{sky}}$ and volume fraction $f_{\text{vol}}$:
\begin{equation}
\Delta \ell \approx \frac{1}{f_{\text{sky}}}, \quad \Delta n \approx \frac{f_{\text{sky}}}{f_{\text{vol}}}.
\end{equation}
For our survey geometry, this yields bin widths of $\Delta n = 1$ and $\Delta \ell \in \{1, 2\}$, depending on the target angular resolution.

In the limit of statistical homogeneity and isotropy, the SFB power spectrum is related to the traditional 3D power spectrum $P(k)$ by:
\begin{equation} \label{eq:ClPk}
C_{\ell}(k, k') = \delta^{D}(k - k') P(k).
\end{equation}
Incorporating observational effects—including the radial selection function $\phi(r)$, the linear growth factor $D(r)$, scale-dependent bias $b(k, r)$, and redshift-space distortions (RSD)—the SFB power spectrum can be modeled via the Limber approximation \citep{2021PhRvD.104l3548G}:
\begin{align} \label{eq:limb}
  C_{\ell}(k, k') &\approx P(k) e^{-\sigma_u^2 k^2} \delta^{D}(k - k')\\ \notag
               & \times \left[ \phi(r) D(r) b(k, r) \mathcal{F}_{\text{RSD}}(\ell, k, r) \right]^2_{r = \frac{\ell + 1/2}{k}},
\end{align}
where $e^{-\sigma_u^2 k^2}$ accounts for the Gaussian Finger-of-God (FoG) effect with $\sigma_u$ being the pairwise velocity dispersion in units of length, and $\mathcal{F}_{\text{RSD}}$ is the RSD correction factor. This highlights the capacity of the SFB framework to encapsulate complex cosmological and astrophysical signatures—including FoG effects, redshift evolution, and the Kaiser effect—without necessitating the flat-sky approximation.

In the plane-parallel limit, the SFB power spectrum can be approximately mapped to the clustering wedge $P(k, \mu)$ \citep{2025PhRvD.112f3518W}:
\begin{equation} \label{eq:ClPkmu}
C_{\ell n n} \approx P\left(k=k_{n \ell}, \mu=\frac{k_{\parallel, n \ell}}{k_{n \ell}}, r=r_{\text{eff},n\ell}\right),
\end{equation}
where the effective distance $r_{\text{eff}, n \ell}$ is defined for each $(n, \ell)$ mode based on the radial selection function and survey boundaries. This mapping allows the SFB analysis to leverage theoretical frameworks developed for Cartesian clustering wedges while maintaining the advantages of a spherical basis.

A primary advantage of the SFB representation is its inherent ability to localize systematic effects. For instance, foreground residuals—characterized by smooth radial profiles—are predominantly confined to the lowest radial modes ($n=1$). This localization enables the selective mitigation of contaminated modes while preserving the cosmological signal at other scales, a feature we exploit in our subsequent analysis.

\section{Results}
\label{sec:result}

\subsection{UWTS Foreground Mitigation}

We first address the dominant astrophysical foregrounds to isolate the cosmological 21\,cm signal. We employ the mPCA method described in Sec.~\ref{subsec:fore}, which involves decomposing the maps into 10 wavelet scales and one coarse component using the UWTS algorithm, followed by independent PCA cleaning of each component. The number of removed foreground modes, $N_{\text{FG}}$, is determined based on the fractional variance explained 
\begin{equation}
R=\frac{\sum_{i=1}^{N_{\text{FG}}}\lambda_i}{ \sum_{i=1}^{N_\nu}\lambda_i},
\end{equation}
where $\lambda_i$ denotes the $i$-th eigenvalue of the PCA decomposition.

\begin{figure}
\centering
\begin{minipage}{0.48\linewidth}
\centering\includegraphics[width=\textwidth]{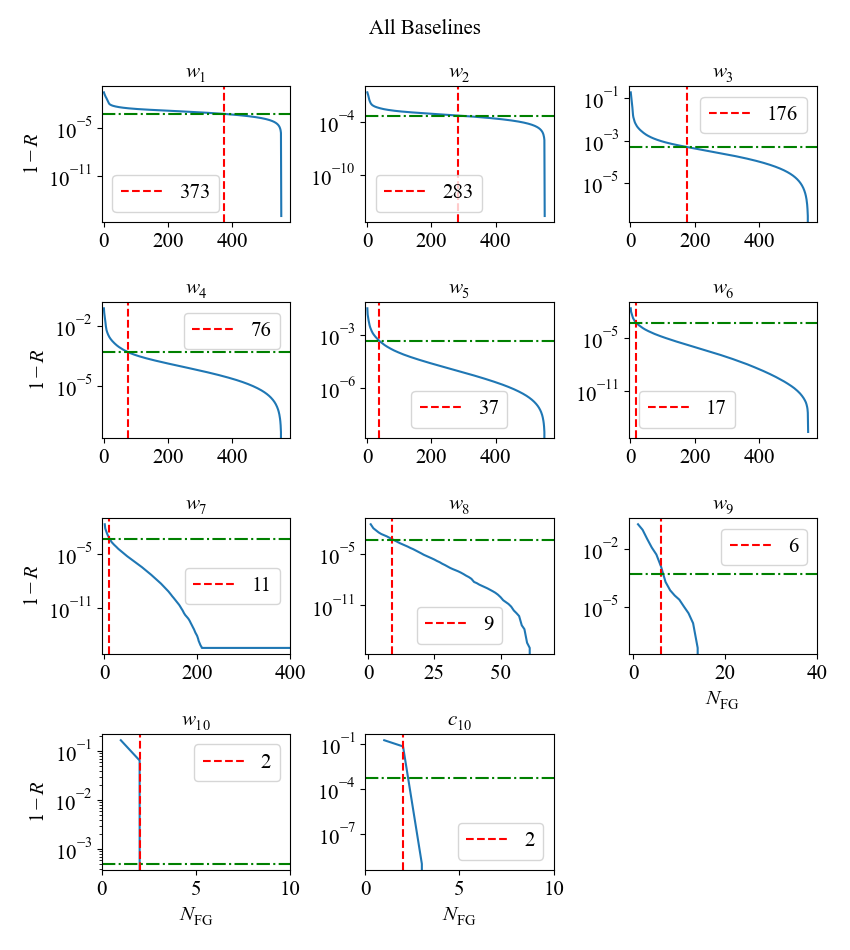}
\end{minipage}
\hfill
\begin{minipage}{0.48\linewidth}
\centering\includegraphics[width=\textwidth]{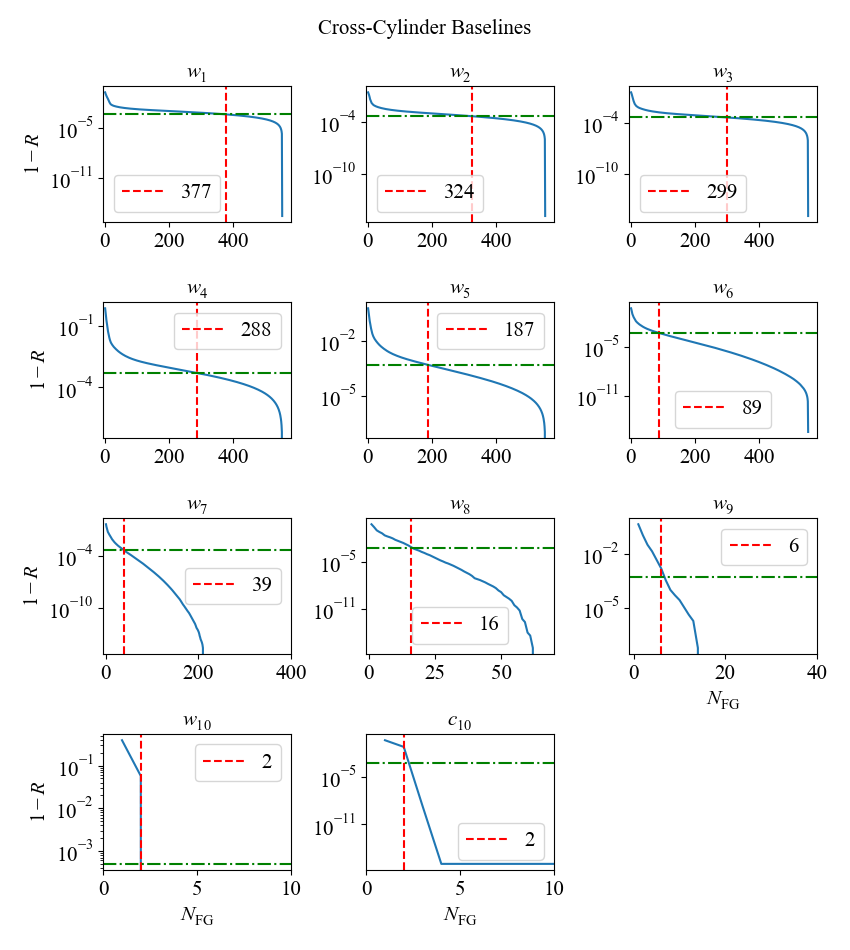}
\end{minipage}
\caption{Fractional retained variance ($1-R$) as a function of PCA mode index for the all-baseline set (left) and the cross-cylinder baseline subset (right). The horizontal dashed-dotted line indicates our chosen cleaning threshold of $0.0005$. The vertical dashed line marks the resulting number of modes ($N_{\rm FG}$) subtracted at each UWTS scale, balancing foreground removal against potential signal loss.
\label{fig:pca}}
\end{figure}

For a given threshold $R$, the required number of modes $N_{\text{FG}}$ varies between the all-baseline set and the cross-cylinder baseline subset. To illustrate this dependence, Figure~\ref{fig:pca} shows the retained variance ($1-R$) as a function of the PCA mode index for both baseline configurations. We adopt a threshold of $1 - R = 0.0005$, indicated by the green dashed-dotted horizontal lines, with the corresponding $N_{\text{FG}}$ marked by red dashed vertical lines. This choice is motivated by the fact that foregrounds are approximately 4--5 orders of magnitude brighter than the expected 21\,cm signal; a value of $0.0005$ provides a conservative balance, avoiding potential over-subtraction associated with more aggressive thresholds (e.g., $10^{-5}$).

The eigenvalue spectra demonstrate a strong scale dependence. For large-scale components ($w_7$--$c_{10}$), the fractional variance explained drops precipitously within the first few modes before reaching a stable plateau, indicating that the foreground power at these scales is highly concentrated in a small number of dominant spectral modes. In contrast, for smaller-scale components ($w_1$--$w_6$), the decline in variance is more gradual, necessitating the removal of a larger number of modes to achieve equivalent suppression. Notably, for the cross-cylinder baseline subset, the required $N_{\text{FG}}$ to reach the same $1-R$ threshold is generally higher than for the all-baseline set. This likely reflects a more complex foreground residual structure or a different noise floor in the cross-cylinder baseline subset.

\begin{figure*}
\centering
\begin{minipage}{0.32\linewidth}
\centering
\text{$w_1$}
\includegraphics[width=\textwidth]{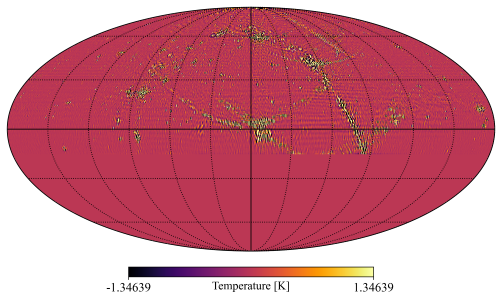}
\end{minipage}
\hfill
\begin{minipage}{0.32\linewidth}
\centering
\text{$w_2$}
\includegraphics[width=\textwidth]{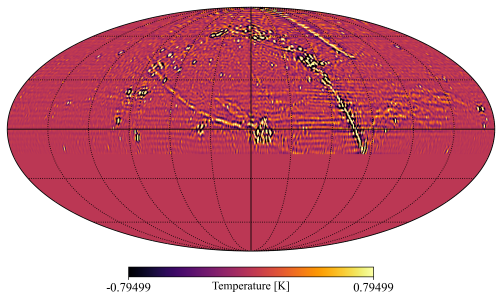}
\end{minipage}
\hfill
\begin{minipage}{0.32\linewidth}
\centering
\text{$w_3$}
\includegraphics[width=\textwidth]{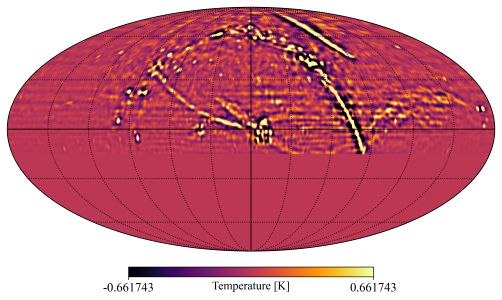}
\end{minipage}
\vfill
\begin{minipage}{0.32\linewidth}
\centering
\text{$w_4$}
\includegraphics[width=\textwidth]{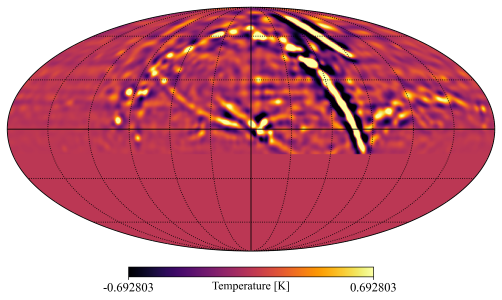}
\end{minipage}
\hfill
\begin{minipage}{0.32\linewidth}
\centering
\text{$w_5$}
\includegraphics[width=\textwidth]{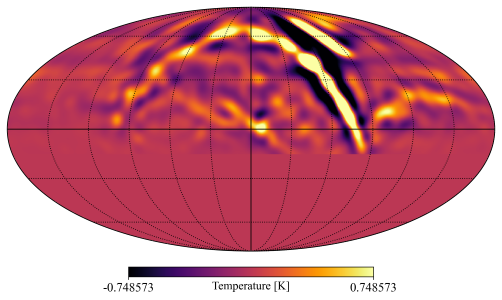}
\end{minipage}
\hfill
\begin{minipage}{0.32\linewidth}
\centering
\text{$w_6$}
\includegraphics[width=\textwidth]{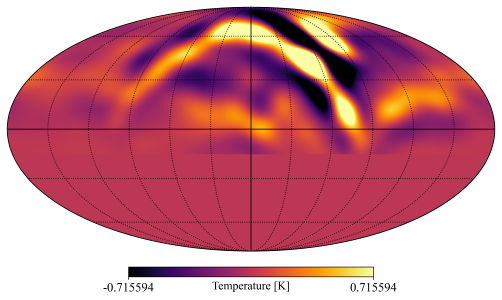}
\end{minipage}
\vfill
\begin{minipage}{0.32\linewidth}
\centering
\text{$w_7$}
\includegraphics[width=\textwidth]{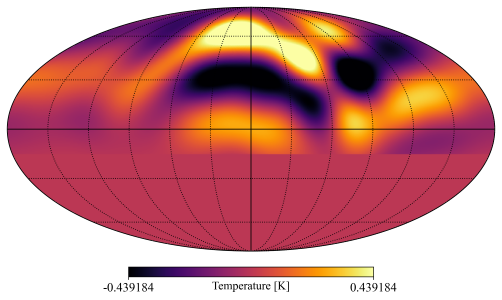}
\end{minipage}
\hfill
\begin{minipage}{0.32\linewidth}
\centering
\text{$w_8$}
\includegraphics[width=\textwidth]{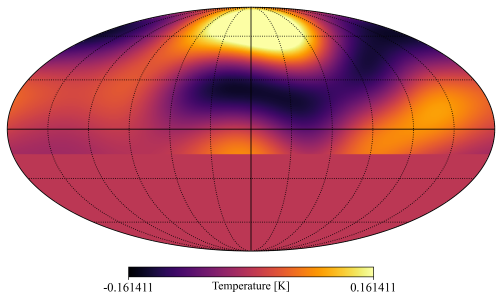}
\end{minipage}
\hfill
\begin{minipage}{0.32\linewidth}
\centering
\text{$w_9$}
\includegraphics[width=\textwidth]{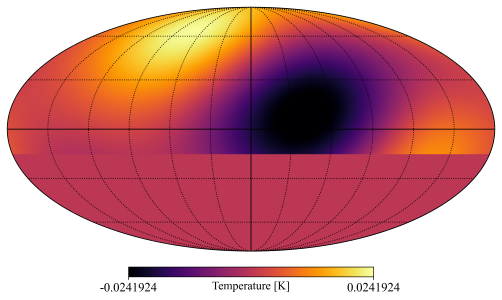}
\end{minipage}
\vfill
\begin{minipage}{0.32\linewidth}
\centering
\text{$w_{10}$}
\includegraphics[width=\textwidth]{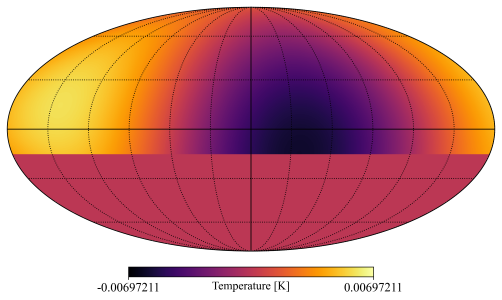}
\end{minipage}
\hfill
\begin{minipage}{0.32\linewidth}
\centering
\text{$c_{10}$}
\includegraphics[width=\textwidth]{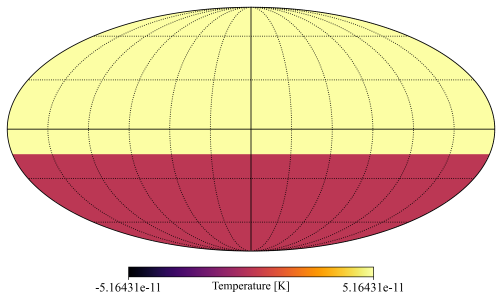}
\end{minipage}
\hfill
\begin{minipage}{0.32\linewidth}
% Empty space to align the two maps above
\end{minipage}
\caption{Multiscale decomposition of the TCPA sky map (all-baseline set) using the isotropic UWTS. The 11 scale maps ($w_1$--$c_{10}$) are displayed at the central frequency of 748\,MHz. The plotting range for each map is restricted to $[-2\sigma, 2\sigma]$ to visualize structures across diverse angular scales.
\label{fig:uwts_a}}
\end{figure*}

Figure~\ref{fig:uwts_a} displays the 11 UWTS scale maps generated using the all-baseline set at the central frequency of 748\,MHz. The maps reveal distinct foreground morphologies across scales: small and medium scales ($w_1$--$w_5$) are dominated by granular structures and point-source-like features, while larger scales ($w_6,\, w_7$) show more diffuse contours. At the coarsest scales ($w_8$--$c_{10}$), the emission transitions into smooth, large-scale asymmetries. The map amplitudes exhibit a systematic decrease towards coarser scales, reflecting the concentration of fluctuation power at higher multipoles. The $c_{10}$ component, which corresponds to the angular monopole or global mean ($\ell \approx 0$), maintains a nearly constant value across the sky and exhibits minimal frequency dependence.

\begin{figure*}
\centering
\begin{minipage}{0.32\linewidth}
\centering
\text{$w_1$}
\includegraphics[width=\textwidth]{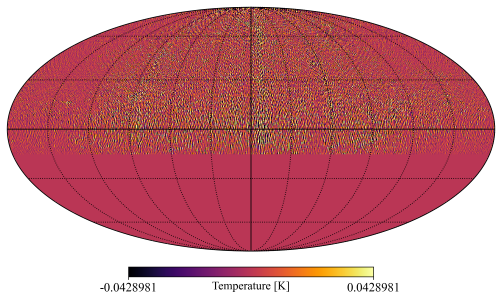}
\end{minipage}
\hfill
\begin{minipage}{0.32\linewidth}
\centering
\text{$w_2$}
\includegraphics[width=\textwidth]{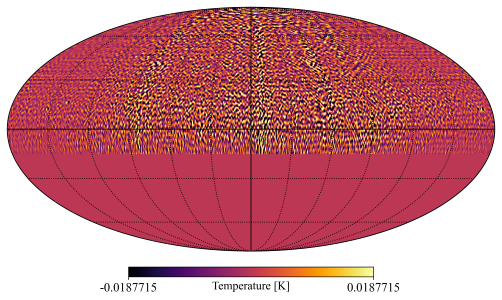}
\end{minipage}
\hfill
\begin{minipage}{0.32\linewidth}
\centering
\text{$w_3$}
\includegraphics[width=\textwidth]{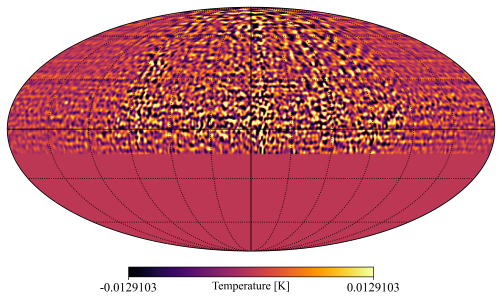}
\end{minipage}
\vfill
\begin{minipage}{0.32\linewidth}
\centering
\text{$w_4$}
\includegraphics[width=\textwidth]{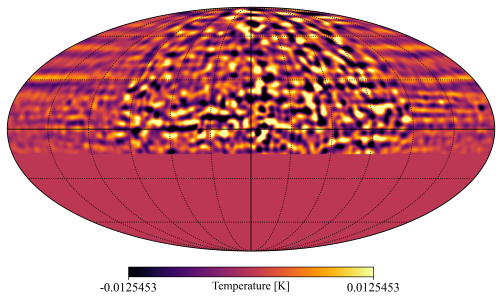}
\end{minipage}
\hfill
\begin{minipage}{0.32\linewidth}
\centering
\text{$w_5$}
\includegraphics[width=\textwidth]{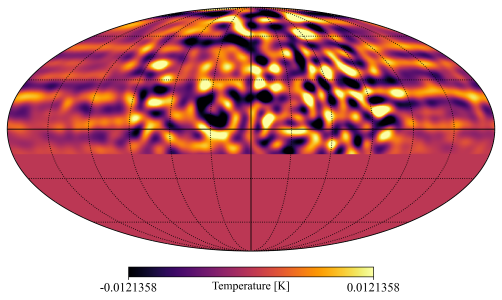}
\end{minipage}
\hfill
\begin{minipage}{0.32\linewidth}
\centering
\text{$w_6$}
\includegraphics[width=\textwidth]{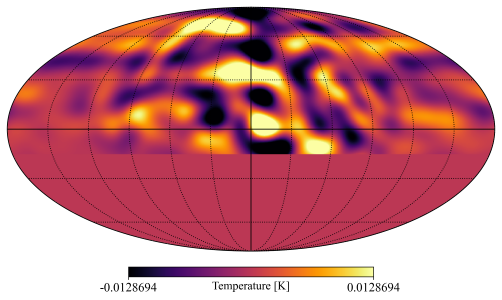}
\end{minipage}
\vfill
\begin{minipage}{0.32\linewidth}
\centering
\text{$w_7$}
\includegraphics[width=\textwidth]{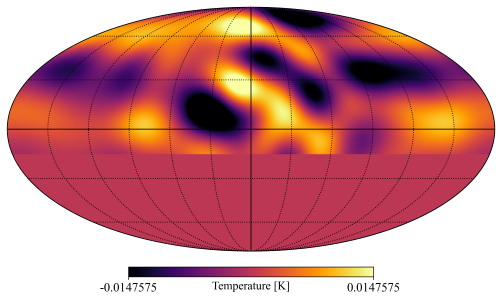}
\end{minipage}
\hfill
\begin{minipage}{0.32\linewidth}
\centering
\text{$w_8$}
\includegraphics[width=\textwidth]{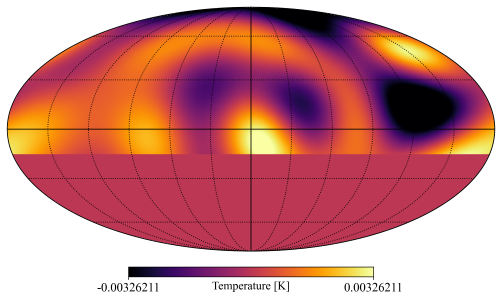}
\end{minipage}
\hfill
\begin{minipage}{0.32\linewidth}
\centering
\text{$w_9$}
\includegraphics[width=\textwidth]{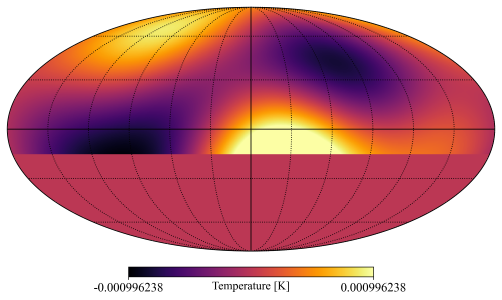}
\end{minipage}
\vfill
\begin{minipage}{0.32\linewidth}
\centering
\text{$w_{10}$}
\includegraphics[width=\textwidth]{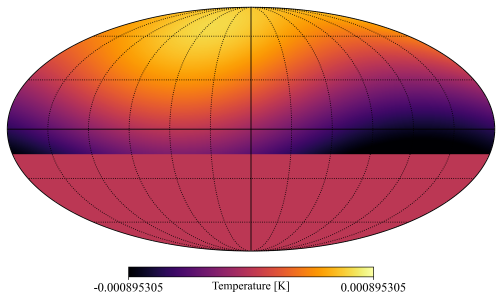}
\end{minipage}
\hfill
\begin{minipage}{0.32\linewidth}
\centering
\text{$c_{10}$}
\includegraphics[width=\textwidth]{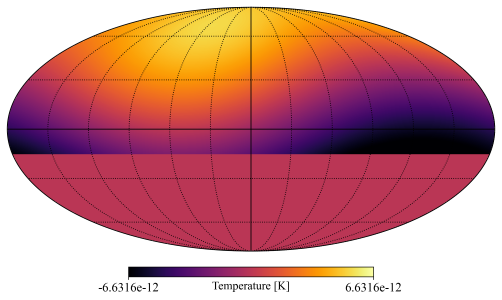}
\end{minipage}
\hfill
\begin{minipage}{0.32\linewidth}
% Empty space to align the two maps above
\end{minipage}
\caption{Residual maps for each of the 11 UWTS scales after independent PCA foreground removal (all-baseline set). The maps are shown at the central frequency (748\,MHz) with a plotting range of $[-2\sigma, 2\sigma]$. The dominant foreground structures seen in Figure~\ref{fig:uwts_a} are significantly suppressed, revealing scale-dependent systematic residuals and potential noise fluctuations.
\label{fig:uwts_res_a}}
\end{figure*}

Figure~\ref{fig:uwts_res_a} presents the residual maps for each of the 11 UWTS scales following independent PCA cleaning using the all-baseline set. The spatial features systematically transition from fine-grained textures at small scales ($w_1$--$w_3$) to intermediate patches ($w_4$--$w_6$), and finally to diffuse asymmetries at the coarsest scales ($w_7$--$c_{10}$). Compared to the original maps (Figure~\ref{fig:uwts_a}), the dominant foreground structures are significantly suppressed, revealing a more irregular morphology with substantially reduced amplitudes. 

We note that the residuals in the $w_1$ to $w_3$ scales exhibit non-uniform features along the longitudinal direction and near the central region, likely reflecting the point spread function (PSF) response and residual point-source contamination. At intermediate scales ($w_4$--$w_6$), a marked increase in variance is observed in sky regions corresponding to daytime observations compared to nighttime data. This underscores the difficulty of removing complex, time-varying artifacts induced by solar transits during the day.

\begin{figure*}
\centering
\begin{minipage}{0.48\linewidth}
\centering
\text{All Baselines}
\includegraphics[width=\textwidth]{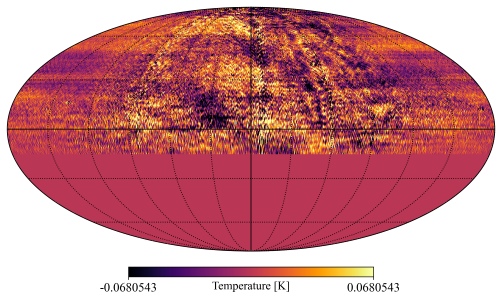}
\end{minipage}
\hfill
\begin{minipage}{0.48\linewidth}
\centering
\text{Cross-Cylinder Baselines}
\includegraphics[width=\textwidth]{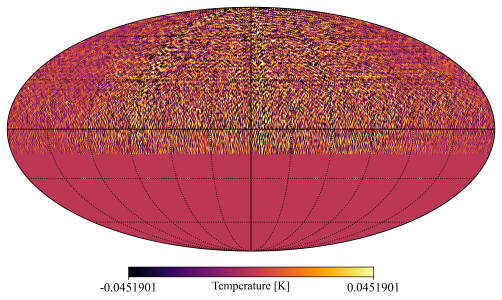}
\end{minipage}
\caption{Final foreground-subtracted residual sky maps at the central frequency for the all-baseline set (left) and the cross-cylinder baseline subset (right). These maps are reconstructed by inverse-transforming the cleaned UWTS scale maps. The plotting range is set to $[-2\sigma, 2\sigma]$ to emphasize the distribution of residual power.
\label{fig:reconst}}
\end{figure*}

The reconstructed foreground-subtracted maps for both the all-baseline set and the cross-cylinder baseline subset are presented in Figure~\ref{fig:reconst}. Comparing these with the raw sky maps (Figure~\ref{fig:map}) highlights a dramatic suppression of astrophysical foregrounds. In the all-baseline reconstruction, intense Galactic plane emission, bright point sources, and extended diffuse structures are effectively mitigated. The resulting maps exhibit a predominantly stochastic texture consistent with instrumental noise and low-level residual systematics, with only minimal large-scale coherent emission remaining near the Galactic plane. Notably, the ``ghost'' aliased artifacts of the Galaxy and the arc-like features from solar transits are significantly suppressed, demonstrating that the mPCA-UWTS framework robustly handles both compact and diffuse foreground contamination.

The cross-cylinder baseline subset yields a more homogeneous, noise-dominated morphology, further reducing the large-scale patchiness seen in the all-baseline results. This underscores the superior foreground isolation of the cross-cylinder baseline subset, though it comes at the cost of reduced sensitivity to the largest angular scales. While the all-baseline set preserves broader spatial information, it requires more stringent control of residual Galactic emission for power spectrum estimation. In both cases, a persistent brightness enhancement in daytime regions remains, reflecting the extreme intensity of solar interference and the ongoing challenge of its complete removal.

\subsection{Power Spectrum for Mock Data}

We utilize the publicly available \texttt{SuperFab} code \citep{2021PhRvD.104l3548G} to compute the SFB power spectrum, which we adapted for 21\,cm intensity mapping analysis. 

\begin{figure}
\centering
\includegraphics[width=0.48\textwidth]{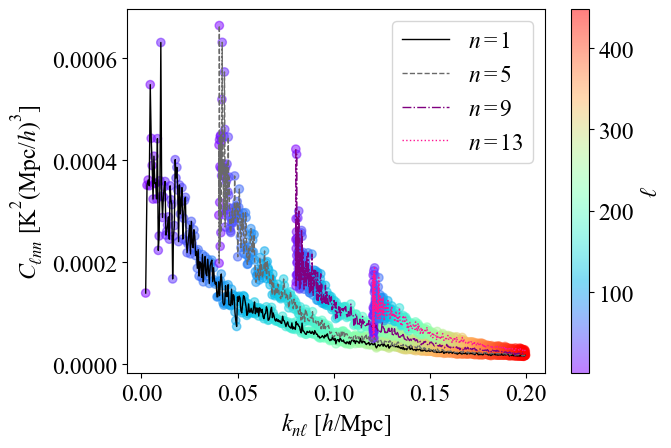}
\caption{Three-dimensional SFB power spectrum of the simulated 21\,cm signal prior to window function deconvolution and bandpower binning for some $n$ values. Individual multipoles $\ell$ are distinguished by colors of markers. The black solid line, gray dashed line, purple dashed-dotted line and pink dotted line represent $n=1,\, 5,\, 9,\, 13$ respectively.
\label{fig:clnn_s_n}}
\end{figure}

\begin{figure*}
\centering
\begin{minipage}{0.48\linewidth}
\centering\includegraphics[width=\textwidth]{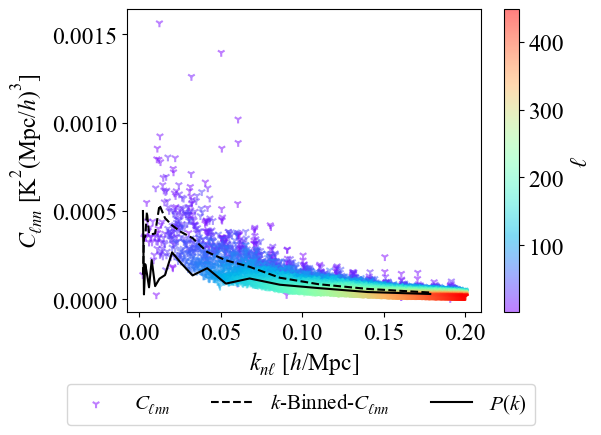}
\end{minipage}
\hfill
\begin{minipage}{0.48\linewidth}
\centering\includegraphics[width=\textwidth]{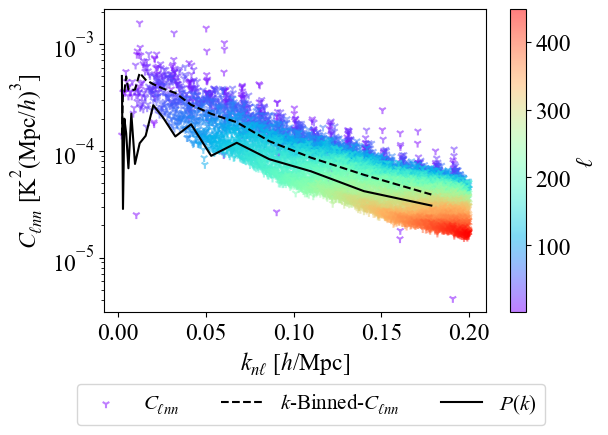}
\end{minipage}
\caption{Three-dimensional SFB power spectrum of the simulated 21\,cm signal prior to window function deconvolution and bandpower binning. Individual multipoles $\ell$ are distinguished by color. The results are presented on both linear (left) and logarithmic (right) scales to illustrate the dynamic range and the characteristic $k$-dependence of the signal.
\label{fig:clnn_s}}
\end{figure*}

To characterize the SFB power spectrum of the 21\,cm signal, we first analyze simulated maps  containing only the cosmological signal. The resulting SFB power spectra $C_{\ell nn}$, prior to window deconvolution or bandpower binning, are shown in Figure~\ref{fig:clnn_s}.
For comparison, we also plot the $k$-binned $C_{\ell nn}$ (black dashed line) and the spherically averaged power spectrum $P(k)$ (black solid line). 
Most $C_{\ell nn}$ values are clustered between $\sim 10^{-5}$ and $\sim 10^{-3}\, \text{K}^2\, \text{Mpc}^3\, h^{-3}$. We exclude the $\ell=0$ modes, as they are sensitive to the global mean and are dominated by numerical noise in our implementation. We also verified that the UWTS decomposition and reconstruction process introduces errors below $\sim 10^{-6}\, \text{K}^2\, \text{Mpc}^3\, h^{-3}$, confirming the signal integrity is preserved. To gain a better understanding of the result, we show $C_{\ell nn}$ for some specific $n$ in Figure \ref{fig:clnn_s_n}, which clearly shows that the characteristic tracks present in $C_{\ell nn}$ are for different $n$s, and the discrete points are due to the discrete nature of $k_{n\ell}$.

The $k$-binned $C_{\ell nn}$ and $P(k)$ display oscillatory patterns at low $k$, with the $k$-binned $C_{\ell nn}$ exhibiting a characteristic peak, for $P(k)$ at $k \approx 0.023\,h\,\text{Mpc}^{-1}$, slightly shifted from the $C_{\ell nn}$ peak at $\approx 0.012\,h\,\text{Mpc}^{-1}$, likely due to window function effects and the discrete nature of the SFB modes. This peak is consistent with the predicted peak in the matter power spectrum. While $P(k)$ follows the overall trend of $C_{\ell nn}$, it exhibits amplitude oscillations due to the uneven binning. At low $k$, $P(k)$ is systematically lower than $C_{\ell nn}$, reflecting the breakdown of the flat-sky approximation.

\subsection{Power Spectrum for Observational Data}

\begin{figure*}
\centering
\begin{minipage}{0.48\linewidth}
\centering
\text{All Baselines}
\includegraphics[width=\textwidth]{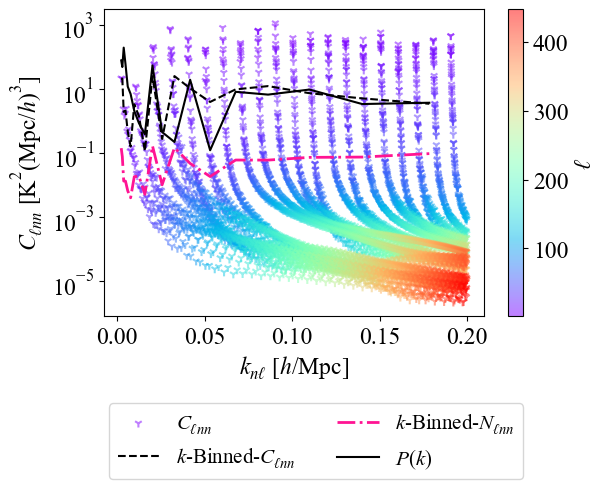}
\end{minipage}
\hfill
\begin{minipage}{0.48\linewidth}
\centering
\text{Cross-Cylinder Baselines}
\includegraphics[width=\textwidth]{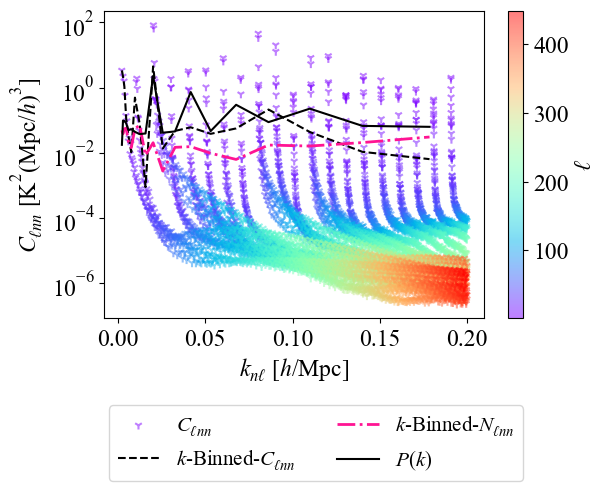}
\end{minipage}
\caption{SFB power spectra for the foreground-subtracted all-baseline set (left) and cross-cylinder baseline subset (right) before window deconvolution and binning. The raw $C_{\ell nn}$ modes (colored markers) exhibit significant scatter due to residual foregrounds and noise. The pink dash-dotted line represents the $k$-binned power spectrum of the processed noise simulations, providing an empirical estimate of the noise floor.
\label{fig:clnn}}
\end{figure*}

Figure~\ref{fig:clnn} displays the SFB power spectrum for the foreground-subtracted observational data. The observational results contrast sharply with the simulations: for the all-baseline set, the observed $C_{\ell nn}$ modes span a dynamic range from $\sim 10^{-6}$ to $10^{3}\, \text{K}^2\, \text{Mpc}^3\, h^{-3}$, reflecting significant residual foregrounds and instrumental noise. In the cross-cylinder baseline subset, the overall amplitude is significantly lower, with most $C_{\ell nn}$ values falling below $10^{-4}\, \text{K}^2\, \text{Mpc}^3\, h^{-3}$. The traditional $P(k)$ estimator yields results comparable in magnitude to the $k$-binned $C_{\ell nn}$ across the entire $k$-range but exhibits prominent oscillations at $k < 0.08\,h\,\text{Mpc}^{-1}$ due to large noise for relatively sparser mode samples in larger scales.

To quantify the noise level and potential pipeline-induced biases, we performed a suite of simulations using the \texttt{tlpipe} package. We generated zero-signal noise visibilities assuming a 20-day observation with a system temperature of 90\,K \citep{2020SCPMA..6329862L}, utilizing the exact TCPA array configuration. These noise realizations were first processed through the map-making pipeline to generate noise maps. To account for the coupling between foregrounds and noise during subtraction, these noise maps were added to the observational sky maps (prior to foreground cleaning), and the combined maps were then passed through the mPCA-UWTS foreground subtraction module. In the mPCA stage, we subtracted the same number of modes per scale ($N_{\text{FG}}$) as used for the primary data (see Figure~\ref{fig:pca}) to ensure the noise residuals correctly reflect any mode-mixing or signal loss effects. The resulting noise power spectra are plotted as pink dash-dotted lines in Figure~\ref{fig:clnn}. This simulation-based approach provides a more robust characterization of the effective noise floor and systematic artifacts than simplified analytical models such as \texttt{21cmSense} \citep{2024JOSS....9.6501M}.

The observed power in the foreground-subtracted data is generally an order of magnitude higher than the noise level, indicating the presence of significant residual foregrounds and systematic artifacts. While a more aggressive PCA cleaning could further reduce these residuals, it would also increase the risk of signal loss. Given the current sensitivity of the Tianlai Cylinder Pathfinder Array and the limited 20-day integration time, we adopt a conservative subtraction strategy. Future work will involve analyzing larger datasets with enhanced foreground mitigation techniques.

\begin{figure*}
\centering
\begin{minipage}{\linewidth}
\centering\includegraphics[width=\textwidth]{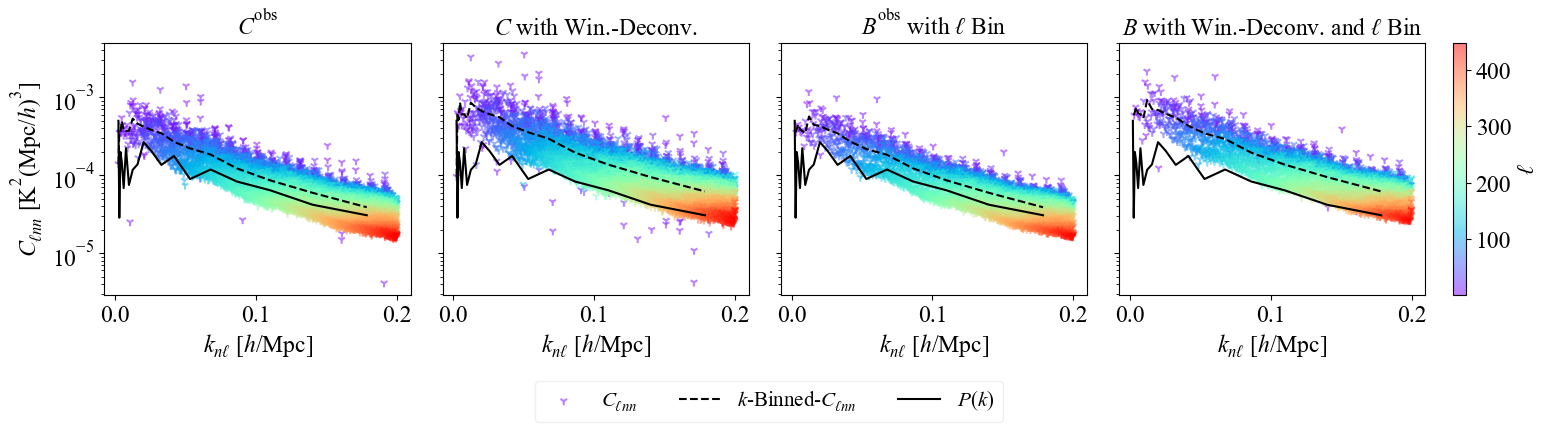}
\end{minipage}
\vfill
\begin{minipage}{\linewidth}
\centering\includegraphics[width=\textwidth]{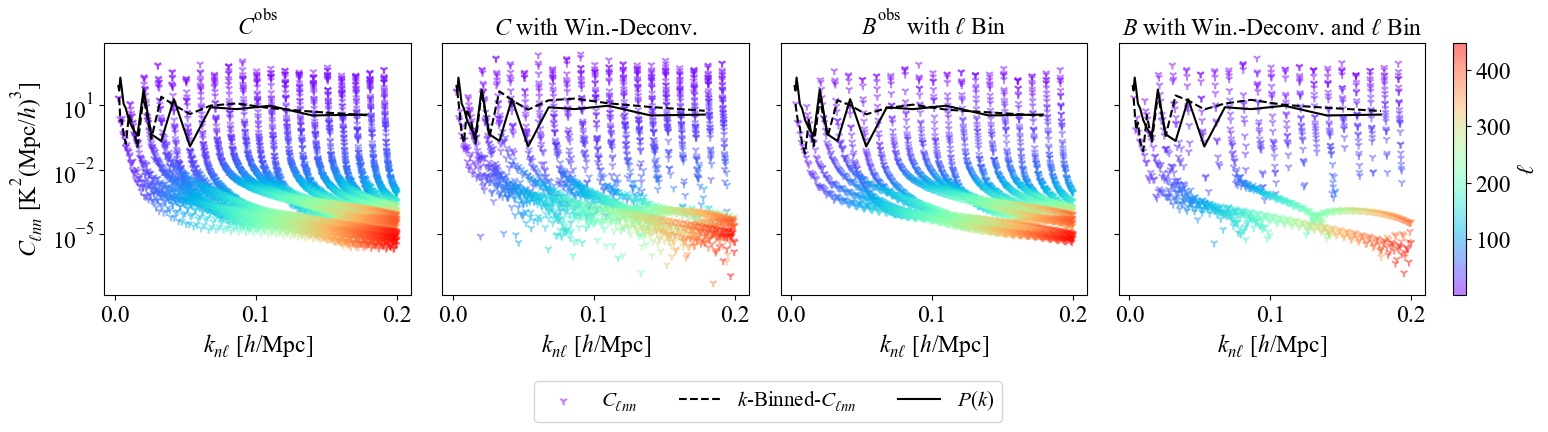}
\end{minipage}
\vfill
\begin{minipage}{\linewidth}
\centering\includegraphics[width=\textwidth]{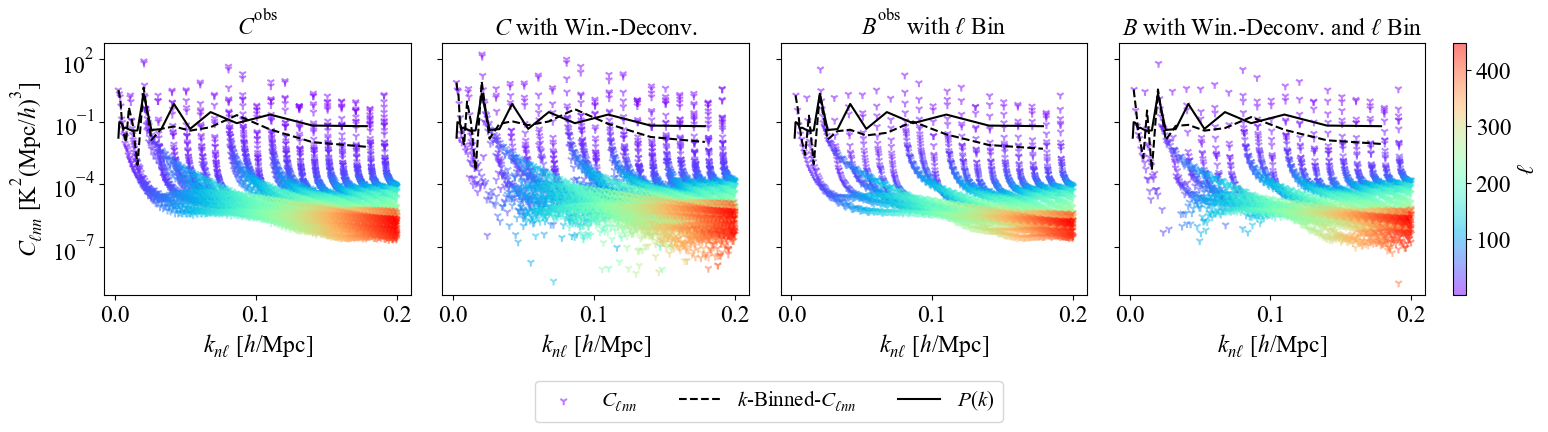}
\end{minipage}
\caption{Impact of window deconvolution and bandpower binning on the SFB power spectrum. \textit{Top}: Simulated 21\,cm signal. \textit{Middle}: Foreground-subtracted all-baseline data. \textit{Bottom}: Foreground-subtracted cross-cylinder subset. In each panel, we compare the raw $C_{\ell nn}$ modes (colored markers), the $k$-binned spectrum (black dashed line), and the spherically averaged power spectrum $P(k)$ (black solid line). Empty bins or excluded data points represent regions with insufficient mode density or negative values.
\label{fig:clnn_vs}}
\end{figure*}

We quantitatively assess the impact of two critical post-processing steps: window function deconvolution and bandpower binning, as defined in Eq.~(\ref{eq:wf}) and Eq.~(\ref{eq:bb}), respectively. Since $\Delta n = 1$ for our survey, no bandpower binning is applied in the radial direction; hereafter, we refer to this process as angular multipole ($\ell$) binning. Their effects are illustrated in Figure~\ref{fig:clnn_vs}. In the simulated data, window deconvolution results in a systematic increase in the $C_{\ell nn}$ amplitude, while $\ell$-binning introduces minimal changes. However, deconvolution also increases the dispersion for low-$\ell$ modes and occasionally yields unphysical negative values (down to $\sim -10^{-3}\, \text{K}^2\, \text{Mpc}^3\, h^{-3}$ for $\ell \le 9$). The deconvolved spectra exhibit a broader distribution, whereas $\ell$-binning leads to a more compact representation. Applying deconvolution after $\ell$-binning mitigates these unphysical values while maintaining the elevated power level and a tighter distribution.

For the observational data, the all-baseline set shows a similar increase in the $k$-binned deconvolved power (black dashed line). However, the dispersion is significantly larger, with substantial unphysical values appearing across a wide range of multipoles, resulting in an irregular spectral shape. In the cross-cylinder baseline subset, window deconvolution also increases the $C_{\ell nn}$ amplitude, though the effect is less pronounced. The occurrence of unphysical values is reduced compared to the all-baseline set, and $\ell$-binning further suppresses these artifacts.

\section{Discussion} 
\label{sec:diss}

The mPCA-UWTS foreground subtraction strategy, unlike traditional methods that rely primarily on frequency-domain correlations, effectively captures both spatial and spectral foreground signatures. This multi-scale approach is essential for wide-field observations like those from the TCPA, where foreground structures exhibit significant spatial heterogeneity. Scale-dependent eigenvalue distributions reveal that large-scale emission ($w_7$--$c_{10}$) is dominated by a few coherent modes, consistent with spectrally smooth Galactic synchrotron emission. In contrast, small-scale fluctuations require more principal components ($N_{\text{FG}}$) to account for their variance, as instrumental noise, point-source confusion, and frequency-dependent beam chromaticity become more prominent and less spectrally correlated. These beam effects introduce fine-grained spectral structures that are not perfectly coherent across the band, posing a challenge for standard PCA techniques.

Notably, the cross-cylinder baseline configuration consistently requires more modes than the all-baseline set at small scales (e.g., 283 vs. 324 for $w_2$). This suggests that the longer baselines sample spatial frequencies where small-scale residuals and noise properties differ significantly, potentially leading to greater mixing between foregrounds and the cosmological signal. Compared to Gaussian Process Regression (GPR), which typically models foregrounds using flexible covariance kernels, our mPCA-UWTS method offers superior computational scalability for large datasets while achieving comparable separation performance. However, GPR may excel in modeling localized spectral features like beam ripples, suggesting that a hybrid approach—using mPCA for dominant foreground removal followed by GPR for residual systematics—could be a promising future direction.

The primary theoretical advantage of the SFB framework lies in its exact treatment of spherical sky geometry, eliminating the need for the flat-sky approximation used in conventional $P(k)$ estimators. Standard Cartesian $P(k)$ methods implicitly assume a fixed line-of-sight (LoS) direction across the survey volume, which breaks down for wide-field observations. In contrast, the SFB basis functions naturally account for sky curvature and the radial variation of the comoving distance. As shown in Figure~\ref{fig:knl}, the discrete radial modes $k_{n\ell}$ form a non-uniform grid that adapts to the survey geometry, providing a more faithful representation of the 3D field than the uniform FFT grid.

Furthermore, the SFB formalism provides a clean separation of angular and radial modes via the quantum numbers $(\ell, n)$, offering diagnostic power that is not achievable with Cartesian Fourier analysis. This decomposition enables the targeted identification and removal of systematic contaminants based on their characteristic scale signatures, avoiding the spectral leakage across modes inherent to Cartesian $k$-space binning. For instance, one can directly excise large-scale radial modes for which systematic effects cannot be reliably mitigated. The behavior of $C_{\ell nn}$—which declines with increasing $\ell$ for fixed $n$ and eventually flattens—reflects a signal increasingly dominated by noise or spectrally uncorrelated residuals at high angular multipoles, motivating differentiated treatment of these modes. The traditional $P(k)$ estimator often exhibits oscillations at low $k$ due to the cylindrical binning scheme, which can obscure real spectral features; the SFB power spectrum is less sensitive to such user-defined binning choices.
The comparison between the all-baseline set and the cross-cylinder baseline subset highlights the fundamental trade-off between sensitivity and systematic control. The all-baseline set retains maximum sensitivity to large-scale modes, capturing the full range of angular scales but requiring more aggressive foreground subtraction. Conversely, the cross-cylinder baseline subset—which excludes intra-cylinder baselines—filters out much of the large-scale diffuse emission, yielding power spectra with amplitudes suppressed by 1--2 orders of magnitude. While this results in a cleaner probe closer to the expected 21\,cm signal amplitude ($z \approx 0.8$--$1.0$), it comes at the cost of reduced sensitivity to the large-scale fluctuations necessary for BAO detection.

Finally, we find that window function deconvolution is essential for unbiased power spectrum recovery, though it introduces stability challenges. In simulations, deconvolution correctly accounts for the fractional sky coverage ($f_{\text{sky}} \approx 62.94\%$) but increases scatter at low $\ell$ and occasionally yields unphysical negative values due to the inversion of ill-conditioned mixing matrices. We demonstrate that applying deconvolution after angular multipole ($\ell$) binning significantly stabilizes the inversion, as binning reduces the effective noise level in each bandpower. In the observational data, however, the presence of strong foreground residuals leads to more frequent unphysical values, particularly in the all-baseline set. The more favorable performance of the cross-cylinder subset confirms that the feasibility of deconvolution depends critically on the quality of the preceding foreground subtraction.

\section{Conclusions}
\label{sec:sum}

In this study, we present the first application of the Spherical Fourier--Bessel (SFB) power spectrum framework to observational 21\,cm intensity mapping data from the Tianlai Cylinder Pathfinder Array (TCPA). To mitigate the dominant astrophysical foregrounds, we implemented a multi-scale subtraction strategy—mPCA-UWTS—which leverages the isotropic Undecimated Wavelet Transform on the Sphere (UWTS) to decompose sky maps into distinct spatial scales, followed by an independent Principal Component Analysis (PCA) within each wavelet domain. This approach effectively isolates complex foreground signatures while preserving the large-scale cosmological signal. Our results demonstrate that the SFB framework provides a mathematically rigorous treatment of the curved sky, offering a robust alternative to Cartesian $P(k)$ estimators for future wide-field 21\,cm surveys.

The primary conclusions of this work are summarized as follows:
\begin{itemize}
    \item The mPCA-UWTS algorithm provides a highly flexible and computationally efficient approach to foreground removal. By performing scale-dependent mode subtraction, the method successfully isolates both diffuse Galactic emission and compact sources, achieving a reconstruction superior to standard spectral-only PCA techniques.
    \item The SFB power spectrum, $C_{\ell nn}$, naturally accounts for the spherical geometry of the sky, eliminating the need for flat-sky approximations. It provides a clean separation of angular ($\ell$) and radial ($n$) modes, thereby avoiding the artificial power-spectrum oscillations often introduced by the cylindrical binning schemes used in Cartesian Fourier analysis.
    \item Baseline selection serves as a critical hardware-level systematic control. We find that the cross-cylinder baseline configuration—which excludes intra-cylinder correlations—yields a power spectrum with significantly lower residual contamination, providing a cleaner (though less sensitive) probe of the $z \approx 0.8$ Universe.
    \item Proper window function deconvolution is essential for an unbiased power spectrum estimation. We demonstrated that a hybrid post-processing strategy—applying deconvolution after angular multipole ($\ell$) binning—stabilizes the matrix inversion process and successfully mitigates the occurrence of unphysical negative power values.
\end{itemize}
These results establish the SFB formalism as a powerful and scalable diagnostic tool for the next generation of 21\,cm intensity mapping experiments. Unlike Cartesian Fourier methods that require uniform gridding and often introduce aliasing artifacts in wide-field surveys, the SFB basis naturally accommodates the spherical geometry with a computational complexity that scales linearly with the number of frequency channels, $O(N_{\text{freq}})$, for each angular multipole $\ell$. This inherent modularity allows for highly parallelized processing across different spatial scales, making it exceptionally well-suited for the massive data volumes expected from the Square Kilometre Array (SKA) and the full Tianlai array. By providing a rigorous treatment of LoS curvature and wide-angle effects within a computationally tractable framework, the SFB approach ensures that the next generation of 21\,cm surveys can fully leverage their large-volume coverage for precision cosmology.

\begin{acknowledgements}
This work is supported by the National SKA Program of China (Nos. 2022SKA0110100 and 2022SKA0110101), the National Natural Science Foundation of China (NSFC) International (Regional) Cooperation and Exchange Project (No. 12361141814), the NSFC (No. 12303004, 12203061.), the Specialized Research Fund for State Key Laboratory of Radio Astronomy and Technology, and the National Astronomical Observatories, Chinese Academy of Science (No. E5ZB0901). The reduction of the Tianlai observational data was carried out at National Supercomputer Center in Tianjin, and the calculations were performed on Tianhe new generation supercomputer.
\end{acknowledgements}

\bibliographystyle{raa}
\bibliography{ms2026-0182}

\label{lastpage}

\end{document}